\def\isarxiv{1}

\ifdefined\isarxiv
\documentclass[11pt]{article}
\else

\fi

\usepackage{tabu}
\usepackage{array}
\usepackage{amsmath}
\usepackage{amsthm}
\usepackage{amssymb}
\usepackage{algorithm}
\usepackage{subfig}
\usepackage{dirtytalk}
\usepackage{color}
\usepackage[english]{babel}
\usepackage{graphicx}
\usepackage{grffile}
\usepackage{wrapfig,epsfig}
\usepackage{epstopdf}
\usepackage{url}
\usepackage{color}
\usepackage{epstopdf}
\usepackage{algpseudocode}
\usepackage[T1]{fontenc}
\usepackage{bbm}
\usepackage{comment}
\usepackage{dsfont}
\usepackage{mathtools}
\usepackage{enumitem}

\usepackage{tikz}
\usepackage{hyperref}  

\hypersetup{
    colorlinks=true,
    citecolor=red,
    linkcolor=blue,
    filecolor=cyan,
    urlcolor=magenta,
}
\usetikzlibrary{arrows}

\ifdefined\isarxiv
\usepackage[margin=1in]{geometry}

\else
\fi

\graphicspath{{./figs/}}

\usepackage[nameinlink,capitalize]{cleveref}
\usepackage{textcomp}
\usepackage{multirow}
\newtheorem{theorem}{Theorem}[section]
\newtheorem{lemma}[theorem]{Lemma}
\newtheorem{definition}[theorem]{Definition}

\newtheorem{corollary}[theorem]{Corollary}

\newtheorem{remark}[theorem]{Remark}

\newtheorem{problem}[theorem]{Problem}

\newcommand{\wh}{\widehat}
\newcommand{\wt}{\widetilde}
\newcommand{\ov}{\overline}
\newcommand{\eps}{\epsilon}

\newcommand{\R}{\mathbb{R}}

\renewcommand{\tilde}{\wt}
\renewcommand{\hat}{\wh}
\renewcommand{\eps}{\epsilon}

\newcommand{\T}{{\cal T}}
\newcommand{\Tmat}{{\cal T}_{\mathrm{mat}}}

\DeclareMathOperator*{\E}{{\mathbb{E}}}

\DeclareMathOperator{\poly}{poly}

\newcommand{\aipe}{\mathsf{AIPE}}

\newcommand{\ade}{\mathsf{ADE}}

\newcommand{\maxip}{\mathsf{MaxIP}}

\newcommand{\lsh}{$\mathsf{LSH}$}
\newcommand{\jl}{\mathsf{JL}}

\newcommand{\ann}{\mathsf{ANN}} 
\newcommand{\fail}{$\mathsf{fail}$}
\newcommand{\FW}{\mathsf{FW}}

\allowdisplaybreaks

\title{Accelerating Frank-Wolfe Algorithm using Low-Dimensional and Adaptive Data Structures}

\ifdefined\isarxiv

\author{
Zhao Song\thanks{\texttt{zsong@adobe.com}. Adobe Research.}
\and
Zhaozhuo Xu\thanks{\texttt{zx22@rice.edu}. Rice University.}
\and 
Yuanyuan Yang\thanks{\texttt{yyangh@cs.washington.edu}. University of Washington.}
\and 
Lichen Zhang\thanks{\texttt{lichenz@uw.edu}. University of Washington.}
}
\else
\author{
}
\fi

\date{}

\begin{document}

\ifdefined\isarxiv
\begin{titlepage}
  \maketitle
  \begin{abstract}
  In this paper, we study the problem of speeding up a type of optimization algorithms called Frank-Wolfe, a conditional gradient method. We develop and employ two novel inner product search data structures, improving the prior fastest algorithm in [Shrivastava, Song and Xu, NeurIPS 2021].
\begin{itemize} 
\item The first data structure uses low-dimensional random projection to reduce the problem to a lower dimension, then uses efficient inner product data structure.  It has preprocessing time $\wt O(nd^{\omega-1}+dn^{1+o(1)})$ and per iteration cost $\wt O(d+n^\rho)$ for small constant $\rho$.
\item The second data structure leverages the recent development in adaptive inner product search data structure that can output estimations to all inner products. It has preprocessing time $\wt O(nd)$ and per iteration cost $\wt O(d+n)$. 
\end{itemize}

The first algorithm improves the state-of-the-art (with preprocessing time $\wt O(d^2n^{1+o(1)})$ and per iteration cost $\wt O(dn^\rho)$) in all cases, while the second one provides an even faster preprocessing time and is suitable when the number of iterations is small.

  \end{abstract}
  \thispagestyle{empty}
\end{titlepage}

\newpage
\else
\maketitle
  \begin{abstract}
  In this paper, we study the problem of speeding up a type of optimization algorithms called Frank-Wolfe, a conditional gradient method. We develop and employ two novel inner product search data structures, improving the prior fastest algorithm in [Shrivastava, Song and Xu, NeurIPS 2021].
\begin{itemize} 
\item The first data structure uses low-dimensional random projection to reduce the problem to a lower dimension, then uses efficient inner product data structure.  It has preprocessing time $\wt O(nd^{\omega-1}+dn^{1+o(1)})$ and per iteration cost $\wt O(d+n^\rho)$ for small constant $\rho$.
\item The second data structure leverages the recent development in adaptive inner product search data structure that can output estimations to all inner products. It has preprocessing time $\wt O(nd)$ and per iteration cost $\wt O(d+n)$. 
\end{itemize}

The first algorithm improves the state-of-the-art (with preprocessing time $\wt O(d^2n^{1+o(1)})$ and per iteration cost $\wt O(dn^\rho)$) in all cases, while the second one provides an even faster preprocessing time and is suitable when the number of iterations is small.

  \end{abstract}
\fi

\section{Introduction}

Using data structure to speed up optimization algorithms has received many success in recent years, and achieved breakthroughs for problems such as linear programming \cite{cls19,lsz19,sy21,dly21,y21,jswz21}, semi-definite programming \cite{jklps20,hjstz22}, sum-of-squares  \cite{jnw22},  non-convex optimization \cite{syz21,bpsw21,szz21}, and discrepancy \cite{sxz22}. Following this trend, we consider the classical optimization algorithm, the Frank-Wolfe ($\FW$) algorithm, and provide algorithms with the best-known running time, improve upon the state-of-the-art~\cite{xss21}.

Frank-Wolfe algorithm is a well-known conditional gradient method with broad applications in machine learning:  In bioinformatics, Frank-Wolfe algorithm is applied for large-scale biological network alignment~\cite{wq16,zxw+19}.  In recommendation systems, Frank-Wolfe algorithm is used for learning user and item representations~\cite{fgm17,ssm17}.
In the intelligent transportation systems, Frank-Wolfe algorithm is the crucial component for traffic assignment~\cite{jtp+94,ml13}.  In robotics, Frank-Wolfe algorithm is recently utilized for policy optimization of agents~\cite{lhy21}.   

Recently, there has been a growing interest among the research community in improving the runtime efficiency of Frank-Wolfe algorithm~\cite{xss21}. To obtain these speedups, these works view the Frank-Wolfe algorithm as an inner product search problem: it preprocesses a bunch of vectors, and at each iteration, it searches for a vector that has the largest inner product with the negative gradient\footnote{We will refer this as the direction search.}, then uses this vector to progress the weight update. The key for attaining faster algorithm is to design data structures that can preprocess the set of vectors, and answers the inner product query quickly, between these vectors and an arbitrary vector. We point out two key observations from this inner product search perspective: 1).\ By modeling the problem as a data structure problem, we focus more on speeding up the \emph{cost per iteration}, instead of total number of iterations. This gives our approach more generality, since any improvement on iteration bounds will lead to improvement to the overall running time. 2).\ Due to this data structure point of view, any new developments in efficient inner product search data structures will lead to improvement to the cost per iteration of our algorithms.

In~\cite{xss21}, they solve the direction search problem via a reduction from nearest neighbor search on unit sphere. Using this observation, they utilize fast locality-sensitive hashing ($\mathsf{LSH}$) data structures~  \cite{im98,diim04,a09,ainr14,ar15,ailrs15,ar16,iw18,r17,alrw17,air18,annrw18,dirw19,anrw21} as a tool for inner product search, and achieve a sublinear cost in the number of vectors to search through. However, their approach has several drawbacks: 1).\ Due to a direct deployment of locality-sensitive hashing, their algorithm has relatively high dependence on the dimension $d$. Specifically, the preprocessing takes time $\wt O(d^2 n^{1+o(1)})$ and each iteration takes time $\wt O(dn^\rho)$. This is particularly troublesome when the vectors-to-search are in the form of self-tensoring~\cite{r10,jzc+16,nmd+19}, inducing large $d$. 2).\ Locality-sensitive hashing data structure is a Monte Carlo data structure. This means that the data structure is \emph{not} robust against an adaptive adversary. During each iteration of the Frank-Wolfe algorithm, one needs to form a negative gradient vector as the input query to the data structure. An adaptive adversary can design the gradient vector using the randomness leaked in prior iterations so that each query will result in failure. This renders the guarantee of the data structure useless in our setting. To resolve this issue, they adapt a net argument, which incurs an extra $d$ factor in both preprocessing, query and space consumption.

To address these concerns, we introduce two data structures that have their own strengths. The first data structure solves the dimension problem by using Johnson-Lindenstrauss transforms~\cite{jl84} to reduce the dimension while preserving the inner products. Then, we feed these low-dimensional vectors into the $\mathsf{LSH}$ data structure. To alleviate the issue of adaptivity, we build two layers of nets: the first layer is for the $\mathsf{JL}$ matrices, which means we'll need to use $\wt O(d)$ many of them. The second layer is for the $\mathsf{LSH}$ data structures. Thanks to the dimension reduction, our second layer of net is built only on dimension of $O(\log n)$, and it suffices to use $\wt O(1)$ $\mathsf{LSH}$s for each $\mathsf{JL}$ matrix. During the query phase, we can simply sample $\wt O(1)$ $\mathsf{JL}$ matrices, and output the best result among them. This approach reduces the dimension on $d$ in preprocessing phase from $d^2$ to $d^{\omega-1}$, and in query phase from $dn^{\rho}$ to $d+n^{\rho}$. This algorithm has particularly small cost per iteration, which gives it much power when one wants a high-precision solution so the number of iterations is large.

The second data structure attacks the problem from the angle of adaptivity. It utilizes a class of data structures for the distance estimation or distance oracle problem~\cite{cn20,cn22,dswz22}. The problem can formulated as follows: suppose there are $n$ points $x_1, \cdots, x_n$ in $\R^d$, one is allowed use $\sim nd$ space and preprocessing time to build a data structure such that for query, it can estimate the distance between $q$ all $x_i$'s under certain norms. These data structures are robust against adaptive adversary, and can output all inner product estimations quickly. To find the vector for direct search, we perform a linear scan over all estimates, and output the best of them. The data structures use random matrices that do not reduce the dimension, but only needs $\wt O(1)$ copies for adaptivity. It has considerably the fastest preprocessing time $dn$, and a slightly slower cost per iteration $d+n$. We remark that, when the number of iterations is small, which is the common case in practice. 

\paragraph{Roadmap.}  Section~\ref{sec:related} reviews the related works on Frank-Wolfe algorithms and how to use adaptive data structures in optimization tasks. In Section~\ref{sec:alg_intro} proposes our algorithm and presents the main result showing its convergence rate. In Section~\ref{sec:prof_overview}, we present the proof of our main theorems. In Section~\ref{sec:conclude}, we present the conclusions of this paper. In Section~\ref{sec:preli}, we present the notations and definitions used in the paper. Section~\ref{sec:data_structure_aipe} showcases how to use $\aipe$ data structure to solve maximum inner product search. Section~\ref{sec:lsh_jl_data_stucture} studies the {\lsh}-JL data structure for $\maxip$. 
In Section~\ref{sec:converge}, we show the convergence rate of our accelerated algorithm, In Section~\ref{sec:herding}, we present the application of our algorithm on Herding problem.

\section{Related work}\label{sec:related}

\subsection{Frank-Wolfe Algorithms}\label{sec:related_fw}

Frank-Wolfe ($\FW$) algorithm~\cite{fw56} is one of the most widely used projection-free optimization methods and has witnessed great success in training machine learning models~\cite{j13,gh15,ahhl17,rsps16,rhsps16}. Every iteration of this optimization algorithm consists of two steps: First, find a vector within a region that has the maximum inner product with the negative gradient. Then, update the current weight by this vector. More precisely, given the learning rate $\eta$, a convex loss function $f:\R^d\rightarrow \R$ with respect to a set $S$, and the initialization $w_0$, the optimization produced by Frank-Wolfe algorithm is as follows:
 \begin{align*}
    s^{*} \leftarrow & ~ \arg\max_{s \in S } \langle s, -\nabla f(w_t) \rangle\\
    w_{t+1} \leftarrow & ~ w_t + \eta_t\cdot (s^*-w_t).
\end{align*}

In this paper, we note that this setting is equivalent to the setting where we want to optimize on the the convex hull of $S$. Here $S$ is  called a finite feasible set. Specifically, we discuss the cost reduction per iteration of Frank-Wolfe algorithm, which has been addressed in~\cite{lktj17,xss21}, as opposed to improving the convergence rate for this algorithm over certain domains~\cite{j13,gh15,ahhl17,l19}. Notably, \cite{xss21} breaks a linear barrier in the running time complexity of Frank-Wolfe iterations by reducing it to an inner product search problem and using hashing-based data structure to speed it up.  However, \cite{xss21} is not optimal due to its quadratic dependence on dimension $d$ in preprocessing phase. In this paper, we further improve their results.

\subsection{Adaptive Data Structures for Optimization}

Recent years, there's a growing trend of applying data structures for optimization problems in large scale machine learning~\cite{cxs19,ll19a,cmfgts20,clp+21,xcl+21,xss21,ssx21,sxz22}. One major challenge that almost all such data structures need to tackle is the robustness against adaptive queries. Unlike the standard similarity search regime, the queries for such data structures in each step are adaptive. For instance, the weight vector for current step is dependent on the previous step in gradient descent. To handle this, \cite{xss21} and \cite{ssx21} introduced a net argument to quantize the queries so that the adaptive queries become independent queries. However, this simple technique seems sub-optimal. A more deep white-box design of the data structure should be proposed for efficient optimization.

\section{Faster $\FW$ Algorithms with Faster Data Structures}\label{sec:alg_intro}

We formulate the problem and define some useful notations in this section, then present an overview of the data structures we will be using.

\subsection{Problem Formulation}\label{sec:formulation}

We start with the notations that will be used throughout the paper.

\paragraph{Notations.} Given a positive integer $n$, we denote $[n]$ as the integer set $\{1,2,\ldots, n\}$. Given a vector $x \in \R^{d}$, we denote $\|x\|_2 := \sqrt{\sum_{i \in [n]} x_i^2}$ as its $\ell_2$ norm. A convex function $g$ has the following rule: $g(a)\geq g(b)+\langle \nabla g(b),a-b \rangle$. A $\beta$-smooth function $g$ satisfies $g(b)\leq g(a)+\langle \nabla g(a),b-x \rangle+\frac{\beta}{2}\| b-a \|^2_2$. 

Given a set $\mathcal{K}=\{x_1, \cdots, x_n\} \subset \R^d$. We use $\mathcal{B}(\mathcal{K})$ (convex hull) to denote the set of all convex combinations of $x_i$, i.e, $\mathcal{B}(\mathcal{K}) = \{ y~| y= \sum_{i\in[n]} \lambda_i x_i, s.t.~\lambda \in \R_{\geq 0}^n,~ \sum_{i=1}^n \lambda_i = 1\}$. We denote $D_{\max}$ as the diameter of ${\cal B} (\mathcal{K})$, i.e., $D_{\max} = \max_{(a,b)\in {\cal B}(\mathcal{K})}{\|a-b\|_2}$.

\paragraph{Problem Settings.} Let $S\subset \R^d$ be a set that contains $n$ points and is convex, coupled with a convex and $\beta$-smooth function $f:\R^d\rightarrow \R$. Our goal is to obtain a vector $w^*\in {\cal B(S)}$ such that $w^*=\arg\min_{w\in {\cal B(S)}}~f(w)$. We particularly consider the case for large $d$, or even $d\geq n$. This means it is not enough to design algorithm with good dependence on $n$ but poor dependence on $d$.

\begin{problem}[]\label{prob:task}
Let $S\subset \R^d$ be a set that contains $n$ points. Let the diameter of ${\cal B}$ be  $D_{\max}$. Let $f : \R^d \rightarrow \R$ be a convex and $\beta$-smooth function, and suppose it takes ${\cal T}_f$ time to evaluate the gradient $\nabla f(x)$ for any $x\in \R^d$.

Let $\epsilon\in (0,1)$ be the precision parameter, the goal is to design an iterative algorithm that uses ${\cal S}_{\mathrm{space}}$ space and ${\cal T}_{\mathrm{prep}}$ time in preprocessing, takes $T$ iterations, and ${\cal T}_{\mathrm{cost}}$ time per iteration , starts from a (possibly random) initial point $w_0\in {\cal B}$,
and outputs $w_T \in \R^d$ from ${\cal B}$ such that
\begin{align*}
   f(w_T) - \min_{w\in \cal{B}} f(w) \leq \epsilon, 
\end{align*}
with success probability at least $1-1/\poly(n)$. 
\end{problem}

If we perform Frank-Wolfe algorithm for Problem~\ref{prob:task}, the iteration cost will be dominated by the complexity of finding the maximum inner product item in $S$ for $-\nabla f(w)$. 

\subsection{Our Algorithms}

In this section, we introduce the proposed efficient Frank-Wolfe algorithm.  We start with a reduction from the direction search step to the maximum inner product search data structure problem. As shown in Section~\ref{sec:related_fw}, the direction search procedure searches over the inner product between the negative gradient and the item vectors in the convex set. Our first order of business is a transformation that preserves the inner product between vectors, while putting the vectors onto the unit sphere.

\begin{definition}[\cite{xss21}]\label{def:transform_informal}
Let $\mathcal{K}$ define a convex set. 
Given a function $f : \mathcal{K} \rightarrow \R$, for every $a$ and $b\in \mathcal{K}$, there is a transformation $\psi_1,\psi_2:\R^d\to\mathbb{S}^{d+2}$ that satisfies: $\langle  \psi_1(a) , \psi_2(b) \rangle=C^{-1}\langle b-a,-\nabla f(a) \rangle$ and $\arg\max_{b}\langle  \psi_1(a) , \psi_2(b) \rangle=\arg\max_{b}\langle b-a,-\nabla f(a) \rangle$, where $C$ is some constant.
\end{definition}

The pair of transformations $(\psi_1,\psi_2)$ put points onto unit sphere while preserving the inner product. This enables the duality between inner product and Euclidean distance in standard Euclidean space, and hence we can use data structures that answers distance queries efficiently. As we will show later, these transformations can also be applied to vectors quickly, i.e., in $O(d)$ time. 

To improve upon the result of~\cite{xss21} which has a preprocessing time of $\wt O(d^2 n^{1+o(1)})$ and per iteration cost $\wt O(dn^\rho)$, we use adaptive Johnson-Lindenstrauss matrices as a tool for adaptive inner product estimation, stated as follows:

\begin{theorem}[Adaptive Inner Product Estimation, informal version of Theorem~\ref{thm:aipe}]\label{thm:aipe_informal}
There is a randomized data structure which requires $\tilde{O}(\epsilon^{-2}nd\log(1/\delta))$ space, supports adaptive queries, and provides the following guarantees:

\begin{itemize}
    \item \textsc{Init}$(x_1, x_2, \cdots, x_n, \epsilon, \delta)$ Given data points $\{x_1, x_2, \cdots, x_n\} \subset \mathbb{S}^{d-1}$, an accuracy parameter $\epsilon$ and a failure probability $\delta$ as input. The running time of this operation is  $\wt{O}(\epsilon^{-2} n d \log(1/\delta))$.
    
    \item \textsc{QueryMax}$(q\in\mathbb{S}^{d-1})$: Given a query point $q \in \mathbb{S}^{d-1}$, the \textsc{QueryMax} operation takes $q$ as input and solves the $(1+\epsilon, r)$-$\ann$ data structure problem, where $r\in (0,2)$ satisfies $\min_{x\in X}\|x-q\|_2\leq r$. In addition. The running time of this operation is $\wt O(\epsilon^{-2}(n+d)\log(1/\delta))$.
\end{itemize}
\end{theorem}

The above data structure estimates \emph{all} inner products in an adaptive fashion, then perform a linear scan over all estimates. Note that this data structure has nearly-optimal preprocessing time, since the input size is of $nd$, and has $n+d$ cost per iteration. While the linear dependence on $n$ is fine for small number of iterations, it becomes problematic when one requires exponentially small error and runs the algorithm for a lot of iterations. 

To address this issue, we design a second-type of data structure that further utilizes LSH. It first uses adaptive JL matrices to project the weight vectors and query gradient vector into a small dimension of $O(\epsilon^{-2}\log n)$, then feed these small vectors into LSH data structures. It thus enjoys the small query complexity of LSH, while sacrificing preprocessing time a bit. Overall, this data structures is a strict upgrade from that of~\cite{xss21}.

\begin{theorem}[Informal version of Theorem~\ref{thm:robust_maxip}]\label{thm:robust_maxip_informal}
There is a randomized data structure that requires $\tilde{O}(\epsilon^{-2}dn^{1+o(1)})$ space, supports adaptive queries, and provides the following guarantees:

\begin{itemize}
    \item \textsc{Init}$(x_1, x_2, \cdots, x_n, \epsilon, \delta)$. Given data points $\{x_1, x_2, \cdots, x_n\} \subset \mathbb{S}^{d-1}$. Given an accuracy parameter $\epsilon$. Let $\delta$ be failure probability. The \textsc{Init} operation runs in time $\wt{O}(\epsilon^{-2} (n^{1+o(1)} d+nd^{\omega-1}) \log(1/\delta))$.
    
    \item \textsc{QueryMax}$(q\in\mathbb{S}^{d-1})$: Given a query point $q \in \mathbb{S}^{d-1}$, the \textsc{QueryMax} operation takes $q$ as input and solves the $(1+\epsilon, r)$-$\ann$ data structure problem, where $r\in (0,2)$ satisfies $\min_{x\in X}\|x-q\|_2\leq r$. The running time of this operation is $\wt O(\epsilon^{-2}(d+n^\rho)\log(1/\delta))$. We remark that $\rho$ is some parameter satisfying that $\rho\in (0,1)$.
\end{itemize}

\end{theorem}

Our data structure improves upon the data structure used in~\cite{xss21} on all fronts: by projecting vectors into small dimensions, it improves the initialization from $d^2n^{1+o(1)}$ to $dn^{1+o(1)}+nd^{\omega-1}$, where $\omega\approx 2.37$ is the current matrix multiplication exponent~\cite{w12,aw21}. Many researchers believe that $\omega=2$, and in that case, our data structure achieves an almost linear preprocessing time in terms of input size $nd$. It also decouples the query time from $dn^\rho$ in~\cite{xss21} to $d+n^\rho$. This small dependence on $n$ in query time makes this data structure suitable when number of iterations of the algorithm is large.

With these data structures, we develop the meta algorithm for accelerating Frank-Wolfe as follows: first apply the transformations $\psi_2$ to the set of weights, then feed these vectors into our data structure. For each iteration, we compute the gradient, and query the data structure to find the vector with approximately largest $\mathsf{IP}$ that has negative gradient, then update the weight. We provide a sketch of the algorithm below, with the data structure being ${\sf JLT}+$\lsh. 

\begin{algorithm}[!ht]\caption{Accelerated Frank-Wolfe for Problem~\ref{prob:task}}\label{alg:frank_wolfe_jl_informal}
\begin{algorithmic}[1]
\State {\bf data structure} \textsc{LSH-Table}
    \State \hspace{4mm} \textsc{Preprocess}($S \subset \R^d$, $c\in (0,1)$) 
    \State \hspace{4mm} \textsc{Query}($x \in \R^d$, $r\in(0,1)$) 
\State {\bf end data structure}
\State
\Procedure{Accelerate-Frank-Wolfe}{$S \subset \R^d$, $k\in \mathbb{N}_{+} $, $m\in \mathbb{N}_{+} $, $c\in (0,1)$, $r\in(0,1)$} \Comment{ Theorem~\ref{thm:frank_wolfe_aipe_informal}} 
\State Construct $\psi_1, \psi_2 : \R^d \rightarrow \R^{d+3}$ as in Definition~\ref{def:transform_informal}
\State For $j\in[k]$, let $R_j: \R^{d+3}\rightarrow \R^m$ denote independent JL transforms 

\State {\bf static} \textsc{LSH-Table} $\textsc{lt}_1,\cdots,\textsc{lt}_k$
\For{{ $j=1 \to k$}}
\State $\textsc{lt}_j$.\textsc{Preprocess}($R_j(\psi_2(S)),c$)
\EndFor

\State Start with $w_0 \in {\cal B}(S)$.
\State $T\leftarrow O(c^{-2}\epsilon^{-1}\beta D_{\max}^2)$
\For{ $t=1 \to T-1$}

    \State Sample $j_1,\cdots, j_l$ with replacement from $[k]$
    \For{$i\in[l]$}
    \State $s^* \leftarrow \textsc{lt}_{j_i}.\textsc{Query}(R_{j_i}\psi_1(w_t),r)$ 
    \If{$\langle s^*, \psi_1(w_t)\rangle \geq cr$} \Comment{Find the first vector with desired inner product.}
    \State \textbf{break}
    \EndIf
    \EndFor

    \State $w_{t+1} \leftarrow w_t + \eta_t\cdot (s^*-w_t)$
\EndFor
\State \Return $w_{T}$
\EndProcedure
\end{algorithmic}
\end{algorithm}

We also put a template of the algorithm with the adaptive inner product estimation data structure, implemented via adaptive $\mathsf{JLT}$, as follows.

\begin{algorithm}[!ht]\caption{Accelerated Frank-Wolfe for Problem~\ref{prob:task}}\label{alg:frank_wolfe_aipe_informal}
\begin{algorithmic}[1]
\State {\bf data structure} \textsc{AIPE}
\Comment{Theorem~\ref{thm:frank_wolfe_aipe_informal}}
    \State \hspace{4mm} \textsc{Preprocess}($S \subset \R^d$, $c\in (0,1)$) 
    \State \hspace{4mm} \textsc{Query}($x \in \R^d$, $r\in(0,1)$) 
\State {\bf end data structure}
\State
\Procedure{Accelerated-Frank-Wolfe}{$S \subset \R^d$,  $c\in (0,1)$, $r\in(0,1)$} \Comment{Choice 2 of Theorem~\ref{thm:frank_wolfe_aipe_informal}} 
\State Construct $\psi_1, \psi_2 : \R^d \rightarrow \R^{d+1}$ as Definition~\ref{def:transform_informal}
\State {\bf static} \textsc{AIPE} \textsc{aipe}
\State \textsc{aipe}.\textsc{Preprocess}($\psi_2(S), c$)
\State Begin with $w_0 \in {\cal B}$. 
\State $T\leftarrow O(c^{-2}\epsilon^{-1}\beta D_{\max}^2)$
 
\For{ $j=1 \to T-1$}
    \State $s^* \leftarrow \textsc{aipe}.\textsc{QueryMax}(\psi_1(w_j),r)$ 
    \State $w_{j+1} \leftarrow (1-\eta_j) \cdot w_{j} + \eta_j \cdot s^*$
\EndFor
\State \Return $w_{T}$
\EndProcedure
\end{algorithmic}
\end{algorithm}

\section{Convergence and Runtime Analysis}\label{sec:prof_overview}
We give convergence and runtime analysis for Algorithm~\ref{alg:frank_wolfe_jl_informal} and Algorithm~\ref{alg:frank_wolfe_aipe_informal}. Then, we compare these results with the original algorithms and \cite{xss21} in Table~\ref{tab:main_compare}. We show that, our algorithm further optimizes the  iteration cost which is sublinear in the size of vertices of the convex hull, while maintaining the same number of iterations towards convergence, with linear preprocessing time.

\begin{table}[h]
    \centering
    \begin{tabular}{|l|l|l|l|l|l|l|} \hline
         {\bf References/Statements} & {\bf Space Storage} & {\bf Preprocessing Time} & {\bf Cost per iter}  \\ \hline \hline
        \cite{j13} & $dn$ & 0 & $dn+{\cal T}_{f}$  \\ \hline
       \cite{xss21} &$d n^{1+o(1)}+d^2 n$ & $d^2n^{1+o(1)}$   & $  dn^{\rho} +{\cal T}_{f}$ \\ \hline 
       Choice 1, Theorem~\ref{thm:frank_wolfe_aipe_informal} & $d n^{1+o(1)}$ & $\Tmat(d, d, n)+dn^{1+o(1)}$ & $d+n^{\rho} +{\cal T}_{f}$ \\ \hline 
        Choice 2, Theorem~\ref{thm:frank_wolfe_aipe_informal}  & $dn$ & $ dn $   & $  d+n +{\cal T}_{f}$ \\ \hline \hline
        \cite{blo12} & $dn$ & 0   & $dn$  \\ \hline
        \cite{xss21}  & $dn^{1+o(1)} + d^2 n$  & $d^2 n^{1+o(1)}$   & $  dn^{\rho} $ \\ \hline 
        Choice 1, Theorem~\ref{thm:herding_jl_informal}  & $dn^{1+o(1)}$ & $\Tmat(d, d, n)+d n^{1+o(1)}$ & $  d+n^{\rho} $ \\ \hline 
        Choice 2, Theorem~\ref{thm:herding_jl_informal} & $dn$ & $dn$  & $  d+n $ \\ \hline \hline
    \end{tabular}
    \vspace{4mm}
    \caption{Comparison between our accelerated $\FW$ algorithm and 1).\: the original $\FW$ algorithm
    ~\cite{j13}, 2).\ the algorithm in \cite{xss21}. For simplicity of representation, we ignore the big-Oh notation in the table. Moreover, we also compare a case of $\FW$ algorithm for Herding problem (see Problem~\ref{prob:herding}) with the original algorithm as shown in \cite{blo12}. 
    Let $\Tmat(x,y,z)$ be the time of matrix multiplication of an $x \times y$ matrix with another $y \times z$ matrix. 
     We remark that $n^{o(1)}<n^{c}$ for $c>0$. $\rho \in (0,1)$ is the approximation ratio to maximum inner product search in {\lsh}. The  failure probability of our algorithm is at most $1/\poly(n)$. $\beta$ defines the smoothness of a function. $D_{\max}$ defines the maximum diameter of a convex hull.  
    }
    \label{tab:main_compare}
\end{table}

Next, we introduce the statements of Algorithm~\ref{alg:frank_wolfe_jl_informal} and Algorithm~\ref{alg:frank_wolfe_aipe_informal}.

\begin{theorem}[Convergence result of accelerated Frank-Wolfe]\label{thm:frank_wolfe_aipe_informal}
Let $\psi_1, \psi_2: \R^d \rightarrow \R^k$ be the transformations in Definition~\ref{def:transform_informal}. We define $S \subset \R^d$ be a set of vectors with size $n$, and ${\cal B} \subset \R^d$ be the convex hull of $S$~(see Section~\ref{sec:formulation}). Given a  $\beta$-smooth and convex function $f : \R^d \rightarrow \R$(see Section~\ref{sec:formulation}), suppose it takes ${\cal T}_f$ to evaluate the gradient $\nabla f(x)$ at any point $x\in \R^d$.

Suppose $\epsilon \in (0,0.1)$ is an accuracy parameter. Then, there exists an iterative algorithm (Algorithm~\ref{alg:frank_wolfe_jl_informal} or Algorithm~\ref{alg:frank_wolfe_aipe_informal}) that, with failure probability at most $1/\poly(n)$, starts from a random initialization point $w_0$ from ${\cal B}$, in $T = O( c^{-2} \beta D_{\max}^2 / \epsilon )$ iterations, outputs $w_T$ such that,
\begin{align*}
   f(w_T) - \min_{w\in \cal{B}} f(w) \leq \epsilon, 
\end{align*}
with ${\cal S}_{\mathrm{space}}$ spaces, ${\cal T}_{\mathrm{prep}}$  preprocessing time, and ${\cal T}_{ \mathrm{cost} }$ cost per iteration.

For the running time, we have two choices. For the first choice, we use $\jl$ and \lsh (Algorithm~\ref{alg:frank_wolfe_jl_informal})
\begin{itemize}
    \item ${\cal S}_{\mathrm{space}} = O( d n^{1+o(1)} )$
    \item ${\cal T}_{\mathrm{prep}} = O( \Tmat(d,d,n) + dn^{1+o(1)} )$
    \item ${\cal T}_{\mathrm{cost}} = O(d + n^{\rho} + {\cal T}_f)$
    \item $T = O( c^{-2} \beta D_{\max}^2 / \epsilon )$ 
    \item Note that $c$, $r$ and $\rho$ has the following connection  $\rho:=  \frac{2(1-r)^2}{(1-cr)^2}-\frac{(1-r)^4}{(1-cr)^4}+o(1)$.
\end{itemize}
For the second choice, we use $\aipe$.(Algorithm~\ref{alg:frank_wolfe_aipe_informal})
\begin{itemize}
    \item ${\cal S}_{\mathrm{space}} = \wt{O} (\rho \cdot dn )$
    \item ${\cal T}_{ \mathrm{prep} } = \wt{O}( \rho \cdot dn )$
    \item ${\cal T}_{\mathrm{cost}} = O( \rho \cdot ( d + n ) + {\cal T}_f)$
    \item  $T = O(  c^{-2} \beta D_{\max}^2 / \epsilon  )$ 
    \item Note that $c$, $r$ and $\rho$ has the following connection $\rho:= \frac{(1-\frac{1}{r})^2}{c^2}$.
\end{itemize}
\end{theorem}

Next, we present the problem formulation and results for the Herding algorithm, which has broad applications in kernel method~\cite{cws12}. We extend our results to the Herding algorithm by the equivalence between the Herding algorithm and a CGM with a mean-square error function~\cite{blo12}. The formal statements are as follows: 

\begin{problem}[Herding]\label{prob:herding}
Let ${\cal X}\subset \R^d$ denote a dataset and $\Phi: \R^d \rightarrow \R^k$ denote a linearized kernel transform.   Let $D_{\max}$ be the maximum diameter of ${\cal B}(\Phi({\cal X}))$. For a distribution $P$ defined on ${\cal X}$, we write $\mu= \E_{x\sim P}[\Phi(x)]$. Let $\rho\in(0,1)$, for any error parameter $\epsilon$, the goal is to design an iterative algorithm that uses ${\cal S}_{\mathrm{space}}$ space and ${\cal T}_{\mathrm{prep}}$ time in pre-processing, and ${\cal T}_{\mathrm{cost}}$ computation in each iteration, begins with a random initialization weight $w_0\in {\cal B}(\Phi({\cal X}))$,
and generate $w_T \in \R^k$ from ${\cal B}(\Phi({\cal X}))$ after $T$ iterations such that
\begin{align*}
   \frac{1}{2} \|w_T-\mu\|_2^2  \leq \min_{w\in \cal{B}} \frac{1}{2} \|w-\mu\|_2^2+\epsilon, 
\end{align*}
is holding with probability at least $1-1/\poly(n)$.
\end{problem}

We remark that the Herding problem defined in Problem~\ref{prob:herding} is a special case for Problem~\ref{prob:task} with $1$-smooth function. So Theorem~\ref{thm:frank_wolfe_aipe_informal} naturally implies the following.

\begin{theorem}[Accelerated Herding algorithm]\label{thm:herding_jl_informal}
For any parameters $\epsilon$, we could solve Problem~\ref{prob:herding} with:
\begin{itemize}
    \item ${\cal S}_{\mathrm{space}} = O( dn^{1+o(1)})$, 
    ${\cal T}_{\mathrm{prep}} = O({\cal T}_{\mathrm{mat}} (d,d,n) + d n ^{1 + o(1)})$, 
     $T = O(D_{\max}^2 / \epsilon)$, \\ 
     and ${\cal T}_{\mathrm{cost}} = O( dn^{\rho} )$ using Algorithm~\ref{alg:frank_wolfe_jl_informal}.
    \item${\cal S}_{\mathrm{space}} = O( dn)$ 
    , ${\cal T}_{\mathrm{prep}} = O(dn)$
    , $T = O({D_{\max}^2}/{\epsilon})$,
    \\ 
    and ${\cal T}_{\mathrm{cost}} = O( d+n )$, using Algorithm~\ref{alg:frank_wolfe_aipe_informal}.
\end{itemize}
\end{theorem}

\subsection{Proof of Theorem~\ref{thm:frank_wolfe_aipe_informal}}

In this section, we present the proof for our main result. We start with introducing a probabilistic tool from~\cite{xss21}.

\begin{corollary}[\cite{xss21}]\label{coro:proj_maxip_lsh_informal}

 Let $r \in(0,1)$ be parameters. Let $\psi_1, \psi_2: \R^d \rightarrow \R^{d+3}$ be transformations in Definition~\ref{def:transform_informal}, and we let ${\cal T}_{\psi_1}$ and ${\cal T}_{\psi_2}$ denote the time to compute $\psi_1(a)$ and $\psi_2(a)$, respectively. We set the approximation ratio $c \in (0,1)$. Given an set $X \in \R^d$ with size $n$ and satisfies $\psi_2(X) \subset {\cal S}^{k-1}$, there is a randomized data structure that, for a query vector $a \in \R^d$ with $\psi_1(a) \in {\cal S}^{k-1}$, outputs a vector $c \in X$ such that $\langle \psi_1(a) , \psi_2(c) \rangle \geq c \cdot \max_{b\in X}\langle \psi_1(a) , \psi_2(b) \rangle$ holds with probability $0.99$. 
 This data structures takes
 \begin{itemize} 
 \item $O(dn^{1+o(1)}+{\cal T}_{\psi_2}n)$ in initialization, 
 \item uses space $O(n^{1+o(1)} + d n)$, 
 \item and takes $O(d\cdot n^{\rho}+{\cal T}_{\psi_1})$ query time,
 \end{itemize}
 where $\rho:=  \frac{2(1-r)^2}{(1-cr)^2}-\frac{(1-r)^4}{(1-cr)^4}+o(1)$. 

\end{corollary}

Next, we give the proof of Theorem~\ref{thm:frank_wolfe_aipe_informal}.
\begin{proof}
We first give an intuition of the proof. Consider the following algorithm: At $t$-th iteration, given current weight vector $w_t$, the algorithm starts with an initial guess $r$, then there are two possible cases: (1) $r>\max_{s\in S }\langle \psi_2(s), \psi_1(w_t) \rangle$ or, (2) $r \leq \max_{s\in S }\langle \psi_2(s), \psi_1(w_t) \rangle$.

Note that the algorithm can only proceed if the $r \leq \max_{s\in S }\langle \psi_2(s), \psi_1(w_t) \rangle$, so if we encounter the first case, we can simply divide $r$ by 2 and use $r/2$ as a new search parameter.

Suppose that after $\log(r/\epsilon)$ iterations, we are still in the first case, i.e.,

\begin{align*}
    r/2^{\log_2{r/\epsilon}} \geq & ~ \max_{s\in S }\langle \psi_2(s), \psi_1(w_t) \rangle,
\end{align*}
then, it follows that $\max_{s\in S }\langle \psi_2(s), \psi_1(w_t) \rangle \leq \epsilon$. 
In other words, the algorithm has already converged to the desired solution in this case.

{\bf Convergence.} 
Next, we present the formal proof of convergence. 

We use $t$ to denote a fixed number of iterations. 

Let us consider two cases: 
\begin{itemize}
    \item {\bf Case 1.} $r > \max_{s\in S }\langle \psi_2(s), \psi_1(w_t) \rangle$;  (All inner products are small, algorithm has converged).
    \item {\bf Case 2.} $r \leq \max_{s\in S }\langle \psi_2(s), \psi_1(w_t) \rangle$. (At least one inner product is large, pick it to make progress).
\end{itemize}

\paragraph{Case 1.}

If we encounter this case, we have:

\begin{align*}
   r \geq & ~ \max_{s\in S }\langle \psi_2(s), \psi_1(w_t)  \rangle \\
   \geq & ~ \langle \psi_2(w^*), \psi_1(w_t)   \rangle \\
   = & ~ C^{-1}\langle w_t - w_*, \nabla f(w_t) \rangle \\
   \geq & ~ C^{-1}(f(w_t) - f(w^*)),
\end{align*}

where the first step follows from case 1's condition , the second step follows from $w^* \in S$ and the definition of convex hull in Section~\ref{sec:formulation}, the 3rd step comes from Definition~\ref{def:transform_informal}, and the last step comes from the convexity of $f$ (see Section~\ref{sec:formulation}).

Thus, if we have $r \leq \frac{\eps}{C}$,  
then we show that
\begin{align*}
    f(w_t) - f(w^*) \leq \epsilon.
\end{align*}
This means the current weight is already $\epsilon$-optimal.

\paragraph{Case 2.}

We first upper bound the term $\langle s^*-w_t,\nabla f(w_t) \rangle$ as

\begin{align}\label{eq:c_approx_intro}
    \langle s^*-w_t,\nabla f(w_t) \rangle 
    =&~ -C\langle \psi_2(s^*), \psi_1(w_t)   \rangle\notag\\
    \leq&~ -c\cdot C\cdot\max_{s\in S }\langle \psi_2(s), \psi_1(w_t)   \rangle  \notag\\
    \leq&~ -c\cdot C \cdot\langle \psi_2(w^*), \psi_1(w_t)   \rangle \notag \\
    =&~c \langle w^*-w_t,\nabla f(w_t) \rangle
\end{align}

where the first step follows from Definition~\ref{def:transform_informal}, the second step follows from  Corollary~\ref{coro:proj_maxip_lsh_informal}, the third step comes from the property of convex hull in Section~\ref{sec:formulation}, and the last step comes from Definition~\ref{def:transform_informal}.

For our convenience, for iteration $t+1$, we denote and upper bound $h_{t+1}$ as:
\begin{align}\label{eq:bound_h_t_intro}
    h_{t+1}&=f(w_{t+1})-f(w^*)\notag\\ 
    &\leq f(w_t)+\eta_t\langle s^*-w_t,\nabla f(w_t) \rangle+\frac{\beta D_{\max}^2}{2}\eta_t^2 -f(w^*)\notag\\ 
    &\leq f(w_t)+c\eta_t\langle w^*-w_t,\nabla f(w_t) \rangle+\frac{\beta D_{\max}^2}{2}\eta_t^2 -f(w^*) \notag\\
     &= (1-c\eta_t)f(w_t)+c\eta_t\left(f(w_t)+\langle w^*-w_t,\nabla f(w_t) \rangle\right)+\frac{\beta D_{\max}^2}{2}\eta_t^2 -f(w^*) \notag\\
     &\leq (1-c\eta_t)f(w_t)+c\eta_t f(w^*)+\frac{\beta D_{\max}^2}{2}\eta_t^2 -f(w^*) \notag\\
     &\leq (1-c\eta_t)f(w_t)-(1-c\eta_t) f(w^*)+\frac{\beta D_{\max}^2}{2}\eta_t^2 \notag\\
     &\leq (1-c\eta_t)h_{t}+\frac{\beta D_{\max}^2}{2}\eta_t^2 
\end{align}

Let $v_t=M_t h_t$,

we have:
\begin{align}
    v_{t+1}-v_t&=M_{t+1}\left( (1-c\eta_t)h_{t}+\frac{\beta D_{\max}^2}{2}\eta_t^2\right) -M_{t} h_{t} \notag \\
    &=\left(M_{t+1}(1-c\eta_t)-M_{t}\right)h_{t}+ \frac{\beta D_{\max}^2}{2}M_{t+1}\eta_t^2
\end{align}

Let $M_t=\frac{t(t+1)}{2}$, $\eta_t=\frac{2}{c(t+2)}$, it is sufficient to show that $M_{t+1}(1-c\eta_t)-M_{t}=0$ and $M_{t+1}\frac{\eta_t^2}{2}=c^{-2}\cdot\frac{(t+1)}{(t+2)}< c^{-2}$.

Next, it is sufficient to upper bound $v_{t+1}-v_t$ as $v_{t+1}-v_t
    <c^{-2}\beta D_{\max}^2$
because $M_{t+1}(1-c\eta_t)-M_{t}=0$ and $M_{t+1}\frac{\eta_t^2}{2}=\frac{t+1}{(t+2)c^2}$. The second step follows from $\frac{t+1}{t+2}<1$.

Next, it is sufficient to upper bound $v_t$ as
\begin{align}
    v_t<c^{-2}t\beta D_{\max}^2
\end{align}
As a result, we have
\begin{align}
     h_t=\frac{v_t}{M_t}<\frac{2\beta D_{\max}^2}{c^{2}(t+1)}
\end{align}

We set $t$ in $O(\frac{\beta D_{\max}^2
}{c^2\epsilon})$ so that $h_t\leq \epsilon$. As a result, we finish the proof.

Next, we introduce the proof of running time.

\paragraph{Choice 1 Result.} We provide a proof for Choice 1's space requirement and running time.

{\bf Space.}
Recall that Choice 1 first uses $\wt O(d)$ independent Johnson-Lindenstrauss matrices, and each matrix is associated with $\wt O(1)$ {\lsh} data structures. Given $d$-dimensional input data of $n$ points, we first compress their dimension into $\wt O(1)$ using Johnson-Lindenstrauss, then store them in downstream {\lsh} data structures. For \emph{one} JL matrix and its corresponding $\wt O(1)$ {\lsh}'s, it takes $\wt O(n^{1+o(1)})$ space. Since there are $\wt O(d)$ JL matrices in total, the total space consumption is $\wt O(dn^{1+o(1)})$.

  {\bf Preprocessing time.}
  
  Note that we can batch the $\wt O(d)$ JL matrices together. Since each of them has dimension $\wt O(1)$, this can be treated as a matrix-matrix multiplication between a matrix that has size $\wt O(d)$ by $d$ and another matrix which has size $d$ by $n$. Multiplying those two matrices together will need $\wt O(\Tmat(d,d,n))$ time. Then, we feed $\wt O(1)$-dimensional data into $\wt O(d)$ independent {\lsh} data structures, by Corollary~\ref{coro:proj_maxip_lsh_informal}, this takes $\wt O(dn^{1+o(1)})$ time.
  
  Hence, the preprocessing takes $\wt O(\Tmat(d,d,n)+dn^{1+o(1)})$ time.

 {\bf Iteration Cost.} 
 
  Compute $\nabla f(w_t)$ is ${\cal T}_{f}$. Moreover, it takes $O(d)$ time to perform $\psi_1(w_t)$ based on Definition~\ref{def:transform_informal}. 
  To realize the query operation, we first sample $\wt O(1)$ JL matrices, apply them to the input point takes $\wt O(d)$ time. We then feed points of dimension $\wt O(1)$ into the $\wt O(1)$ downstream {\lsh} data structures, this takes only $\wt O(n^\rho)$ time to retrieve $s^*$. After receiving $s^*$, we need $O(d)$ time to compute $w_{t+1}$. 
  
  We write the complexity as $\wt O(d+n^\rho+{\cal T}_g)$ for $\rho:=  \frac{2(1-r)^2}{(1-cr)^2}-\frac{(1-r)^4}{(1-cr)^4}+o(1)$.

 \paragraph{Choice 2 Result.}
 
We provide a proof for Choice 2's space requirement and running time.

 {\bf Space.}
 
Using Theorem~\ref{thm:aipe_informal} (the space part), we get the space as ${\cal S}_{\mathrm{space}} = \wt{O}(\eps^{-2} nd \log (1/\delta))$, where $\eps = \frac{c-1}{3 (1-1/r)}$ denotes the accuracy guarantee used in Theorem~\ref{thm:aipe_informal}. Plugging the value of $\eps_0$ gives us the space as ${\cal S}_{\mathrm{space}} = \wt{O}( \rho \cdot nd \log (1/\delta))$, for $\rho:= \frac{(1-\frac{1}{r})^2}{c^2}$.

 {\bf Preprocessing Time.}
 
 Using Theorem~\ref{thm:aipe_informal} (the preprocessing procedure part), and plugging the value of $\eps =\frac{c-1}{3 (1-1/r)}$ as above, we have the preprocessing time $\wt{O}(\rho \cdot n d \log (1/\delta))$, for $\rho:= \frac{(1-\frac{1}{r})^2}{c^2}$.

  {\bf Iteration Cost.}
  
 Using Theorem~\ref{thm:aipe_informal} (the query procedure part), and plugging the value of $\eps =\frac{c-1}{3 (1-1/r)} $ as above, we have the cost per iteration as $\wt{O}(\rho \cdot ( n + d ) \log (1/\delta))$, for $\rho:= \frac{(1-\frac{1}{r})^2}{c^2}$.

 \end{proof}
 
 With this proof, we show that we have two choices to further improve the running time complexity of \cite{xss21}'s algorithm: (1) introduce a near linear preprocessing time with iteration cost $O(d+n)$, (2) reduce the cost in each iteration to $O(d+n^\rho)$ with also a $d$ saving in preprocessing. We highlight the second option as our recommendation as it makes Frank-Wolfe efficient in high-dimensional space.
\section{Conclusion}\label{sec:conclude}
In this paper, we develop novel data structures to improve the running time of Frank-Wolfe and Herding algorithm. By adapting low-dimensional random projections, we beat the state-of-the-art method of~\cite{xss21}. The key to our breakthroughs is a set of low-dimensional random projections that are robust against adaptive adversary. To this end, our data structure has either 1).\ preprocessing time $nd^{\omega-1}$ and query time $dn^\rho$, or 2).\ preprocessing time $nd$ and query time $n+d$. The first one is a direct upgrade to the result in~\cite{xss21}, while the second one provides nearly linear preprocessing time in input size and is efficient for small iterations.

\section{Preliminary}\label{sec:preli}

In this section, we introduce the notations and definitions for our paper. We begin with standard notations in Section~\ref{sec:preli_notation}. Next, Section~\ref{sec:preli_lsh_maxip} introduces the definition of $\ann$ and $\maxip$ problem and the definition for our {\lsh} tool. Next, we provide the optimization settings we studied in this paper in Section~\ref{sec:preli_opt}. Finally, Section~\ref{sec:preli_transform} introduces the efficient transforms that relates  $\maxip$ with $\ann$.

\subsection{Notations}\label{sec:preli_notation}

For a random variable $x$, we use $\Pr[x]$ to represent the probability. We let $\E[x]$ to denote the expectation of a random variable $x$ if it exists. We use $\max\{x,y\}$ to denote the maximum value between $x$ and $y$. We write $\min \{x,y\}$ as the minimum value between $x$ and $y$.  We define $\| a \|_2 := ( \sum_{i=1}^n a_i^2 )^{1/2}$ as the Euclidean($\ell_2$) norm for a vector $a$. Suppose $m$ is a positive integer, we write $[m]:=\{1,2,\cdots,m\}$.

\subsection{\texorpdfstring{{\lsh}}{~} and \texorpdfstring{$\maxip$}{~}}\label{sec:preli_lsh_maxip}

Here, we discuss how to solve the maximum inner product search problem ($\maxip$) using the Approximate  Nearest Neighbor ($\ann$) data structures.

We start with defining the $\ann$ problem. We introduce the standard $\ann$ definitions as below:
\begin{definition}[Approximate Nearest  Neighbor ($\ann$)]\label{def:ann:formal}
Given a set $Y \subset \mathbb{S}^{d-1}$ with size $n$ on a unit sphere. Let $\ov{c}$ and $r$ denote two parameters $\ov{c} >1$ and $r \in (0,2)$. 
The $(\ov{c},r)$-Approximate Nearest Neighbor ($\ann$) solves the following problem: for a query $x \in \mathbb{S}^{d-1}$ that have $\min_{ \wh{y} \in Y}\| \wh{y} - x \|_2 \leq r$, output a $\wt{y} \in Y$ with condition $\| \wt{y} - x \|_2 \leq \ov{c} \cdot r$.

\end{definition}

It is known that locality-sensitive hashing ({\lsh})~\cite{im98,diim04} could be a solution to $\ann$ problem. In this paper, we use {\lsh} data structure (see Indyk and Motwani
~\cite{im98}) with definition as below: 

\begin{definition}[Locality-Sensitive Hashing]
Given the approximation parameter $\ov{c}>1$ and probability parameter $1>p_1 > p_2>0$, we define a function family $\mathcal{H}$ to be $(r,\ov{c} \cdot r,p_1,p_2)$-sensitive
if and only if: given any two vectors $a$ and $b \in \R^d$, a function $h$ that uniformly chosen from $\mathcal{H}$ satisfies:

\begin{itemize}
    \item if $\| a-b\|_2 \leq r$, then $\Pr_{h} [ h(a)=h(b) ] \geq p_1$,
    \item if $ \|a-b\|_2 \geq \ov{c} \cdot r$, then $\Pr_{h} [ h(a)=h(b) ] \leq p_2$.
\end{itemize}
\end{definition}

{\lsh} could preprocess the dataset so that the query for $\ann$ can be handled in time sublinear to the dataset size. We present the formal statement as below:

\begin{theorem}[Theorem 1.2 in~\cite{alrw17}
]
\label{thm:ar17:formal}
Let $\ov{c} > 1$ be a parameter. Let $r \in (0,2)$ be another parameter. The $(\ov{c},r)$-$\ann$ on a unit sphere can be solved by a data structure with query time in $O(d \cdot n^{\rho})$,  preprocessing time in $O(dn^{1+o(1)})$ and space in $O(n^{1+o(1)} + d n)$, where $\rho = \frac{2}{\ov{c}^2} -\frac{1}{\ov{c}^4}+o(1)$.

\end{theorem}

Note that $o(1)$ is equivalent to $O(1/\sqrt{\log n})$. One can obtain better $\rho$ by using {\lsh} in~\cite{ar15} 
given that we pay more time in the initialization stage.  

We provide the statement as follows:

\begin{theorem}[\cite{ar15}]\label{thm:ar15:formal}
Let $\ov{c} > 1$ and $r \in (0,2)$. The $(\ov{c},r)$-$\ann$ on a unit sphere $\mathbb{S}^{d-1}$ can be solved  by a data structure with query time $O(d \cdot n^{\rho})$, space $O(n^{1+\rho} + d n)$ and preprocessing time $O(dn^{1+\rho})$, where $\rho = \frac{1}{2\ov{c}^2-1} +o(1)$.
\end{theorem}

We follow the strategy of using \lsh for $\ann$ that solves the approximate $\maxip$ problem, which is defined below:

\begin{definition}[Approximate $\maxip$, Definition A.4 in \cite{xss21}]\label{def:approximate_maxip}

Given an $n$-vector dataset $Y \subset \mathbb{S}^{d-1}$ and two parameters $c \in (0,1)$ and $\tau \in (0,1)$, $(c,\tau)$-{$\maxip$} defines a task that: for a query $x \in \mathbb{S}^{d-1}$ that satisfies $\max_{y\in Y}\langle x , y \rangle \geq \tau$, outputs a vector $z\in Y$ with a condition that $\langle x , z \rangle \geq c \cdot \max_{y \in Y} \langle x,y \rangle$.

\end{definition}

In optimization, it is essential to solve projected versions of both $\maxip$ and approximate $\maxip$~\cite{xss21}. Here we introduce two versions as below:

\begin{definition}[Projected $\maxip$, Definition A.5 in \cite{xss21}]\label{def:projected_maxip} 
Let $\phi, \psi: \R^d \rightarrow \R^k$ be a pair of functions. We define the $(\phi, \psi)$-$\maxip$ problem for a query $q\in \R^d$ with respect to dataset $X\subseteq \R^d$  as below:
\begin{align*}
    (\phi, \psi)\text{-}\maxip (q,X) := \max_{x \in X} \langle \phi(q),\psi(x) \rangle .
\end{align*}

\end{definition}

\begin{definition}[Projected approximate $\maxip$, Definition A.6 in \cite{xss21}]\label{def:proj_approximate_maxip}
Let $\phi, \psi: \R^d \rightarrow \R^k$ be a pair of functions. Given an dataset $Y \subset \R^d $ with size $n$ that $\psi(Y)$ is on the unit sphere, we define the $(c,\phi, \psi,\tau)$-{$\maxip$} problem as: for the query $x\in \R^d$ such that $\phi(x)\in \mathbb{S}^{d-1}$ and $\max_{y\in Y}\langle \phi(x) , \psi(y) \rangle \geq \tau$, we would like to find a vector $z \in Y$ that have $\langle \phi(x) , \psi(z) \rangle \geq c \cdot (\phi, \psi)\text{-}\maxip (x,Y)$.
\end{definition}

\subsection{Optimization Setting}\label{sec:preli_opt}

In this section, we provide the assumptions to the function we would like to optimize.

\begin{definition}[Convex hull]\label{def:cvx_hull}
We define $\mathcal{B}(A)$ to be the convex hull of a set $A=\{a_i\}_{i\in [n]} \subset \R^d$. Here $\mathcal{B}(A)$ is the union of every vector $y=\sum_{i\in [n]} w_i\cdot a_i$ where $w_i\in [0,1]$ for all $i\in[n]$ and $\sum_{i\in [n]}w_i= 1$. Moreover, we define $D_{\max}$ to be the a value that $\|a-b\|_2\leq D_{\max}$ for all $(a,b)\in {\cal B}(A)$.

\end{definition}

We review smoothness of a function.
\begin{definition}[Smoothness]\label{def:smooth}
We say a function $f$ is $B$-smooth if 
\begin{align*}
f(b)\leq f(a)+\langle \nabla f(a),b-a \rangle+\frac{B}{2} \cdot \| b-a \|^2_2 .
\end{align*}
\end{definition}

We define convexity of a function.
\begin{definition}[Convex]\label{def:convex}
We say a function $f$ is convex if 
\begin{align*}
f(a)\geq f(b)+\langle \nabla f(b),a-b \rangle .
\end{align*}
\end{definition}

In the setting described above, there are some properties we can use:

\begin{corollary}[Corollary A.11 in \cite{xss21}]\label{coro:hull_maxip}

Given a convex hull $\mathcal{B}(A)$ built on set $A=\{a_i\}_{i\in [n]} \subset \R^d$, for a query $q\in \R^d$, if $a^*=\arg\max_{a\in A} q^\top a$, then we have $q^\top b \leq q^\top a^*$ for all $b\in \mathcal{B}(A)$.
\end{corollary}

\begin{lemma}[$\maxip$ condition, Lemma A.12 in \cite{xss21}] \label{lemma:max_ip_condition}
Given a set $S \subset \R^d$ with size $n$ and a convex function $f : \R^d \rightarrow \R$, for a vector $x\in {\cal B}(S)$, we show that
\begin{align*}
\min_{s \in S } \langle \nabla f(x) ,  s-x \rangle  \leq 0, ~~~ \forall x\in \cal{B}(S).
\end{align*}
\end{lemma}

As a condition gradient method, the ${\sf FW}$ algorithm requires to choose step size.

\begin{definition}[Step size]
\label{def:eta_t}
Let $t\in \{0,\ldots,T-1 \}$, we define 
\begin{align*}
    \eta_t := & ~ \frac{2}{c(t+2)},
\end{align*}
where $c\in (0,1]$.
\end{definition}

\subsection{Efficient Transformations}\label{sec:transform}\label{sec:preli_transform}

In this section, we introduce a pair of transforms that preserves the inner product while adjusting the norm.

Given two vectors $a,b\in \R^d$ and a differentiable function $g:\R^d\rightarrow \R$, define two functions $\phi_0, \psi_0$ as

\begin{align}\label{eq:asym_trans_direct}
\phi_0 (a) := &  [\nabla g(a) ^\top, a^\top\nabla g(a)]^\top, ~~~
\psi_0(b) := [ -b^\top,1]^\top.
\end{align}

It is not hard to see that the inner product between $\phi_0(a)$ and $\psi_0(b)$ captures the inner product we wish to estimate:

\begin{align}\label{eq:asym_trans_direct_res}
    \langle b-a,\nabla g(a) \rangle =&~-\langle  \phi_0(a) , \psi_0(b) \rangle,\notag\\
    \arg\min_{b\in Y} \langle b-a,\nabla g(a) \rangle=&~\arg\max_{b\in Y} \langle  \phi_0(a) , \psi_0(b) \rangle
\end{align}

Next, we present another pair of transformations $\phi_1,\psi_1:\R^{d} \rightarrow \R^{d+2}$ that transform the $\maxip$ problem into $\ann$ on unit sphere. Given any $a,b\in \R^d$, we define $\phi_1$ and $\psi_1$ as follows:
\begin{align}\label{eq:asym_trans_mips}
  \phi_1(a) =&~ \begin{bmatrix} (D_{x}^{-1}a)^\top & 0 & \sqrt{1-\|a D_{x}^{-1}\|_2^2}  \end{bmatrix}^\top\notag\\
  \psi_1(b) =&~ \begin{bmatrix} (D_{y}^{-1}b)^\top & \sqrt{1-\|bD_{y}^{-1}\|_2^2} & 0 \end{bmatrix}^\top
\end{align}

where $D_x, D_y$ are radius for the datasets $X$ and $Y$ respectively. Note that the transformations put vectors onto the unit sphere, without changing the inner products. One can easily verify that $\arg\max_{b\in Y} \langle \phi_1(a),\psi_1(b)\rangle=\arg\max_{b\in Y} \langle a,b\rangle$.

Combining Eq.~\eqref{eq:asym_trans_direct} and Eq.~\eqref{eq:asym_trans_mips} together, we can get a final pair of transforms:
1).\ $\phi:\R^d\rightarrow \R^{d+3}$ where $\phi=\phi_1\circ\phi_0$, 2).\ $\psi:\R^d\rightarrow \R^{d+3}$ such that $\psi=\psi_1\circ\psi_0$.

Via transformations $\phi$ and $\psi$, we put points onto unit sphere and preserve the inner product at the same time. This enables us to use fast $\ell_2$ norm estimation data structures to solve inner product search problem.

\subsection{Problem Formulation}\label{sec:alg_formulate}
In this section, we give the formulation of the Frank-Wolfe Problem~\ref{problem:frankwolfe}, and provide a meta-algorithm for it.

\begin{problem}\label{problem:frankwolfe}
\begin{align}
    \min_{w\in {\cal B}(S)}&~ g(w)
\end{align}

In this problem, we assume $g$ is a convex function and is differentiable, and $S$ is a set in $\R^d$ with size $n$, we would like to optimize $w$ in the convex hull of $S$ (see Definition~\ref{def:cvx_hull}), which has a maximum diameter $D_{\max}$.
\end{problem}

We focus on solving the above problem via $\FW$ Algorithm (see Algorithm~\ref{alg:frank_wolfe_formal})

\begin{algorithm}[!ht]\caption{Meta-algorithm for ${\sf FW}$}\label{alg:frank_wolfe_formal}
\begin{algorithmic}[1]
\Procedure{Frank-Wolfe}{$S \subset \R^d$}
\State $T\leftarrow O(\epsilon^{-1}\beta D_{\max}^2)$
\State Start with $w^0 \in {\cal B}$. 
\For{ $t \in [T-1]$}
    \State $s^t \leftarrow \arg\max_{s\in S} \langle -\nabla g(w^t),s \rangle  $\label{line:argmin}
    \State $w^{t+1} \leftarrow w^{t}+\eta_t(s^t-w^t)$ \Comment{$\eta_t$ is defined in Definition~\ref{def:eta_t} with $c=1$.}
\EndFor
\State \Return $w^{T}$
\EndProcedure
\end{algorithmic}
\end{algorithm}

We identify that the dominate computation in Algorithm~\ref{alg:frank_wolfe_formal} is that we need to search all elements in $S$ in each iteration to find the best optimization direction. We achieve this goal via fast data structures.

\section{\texorpdfstring{$\aipe$}{~} Data Structure}\label{sec:data_structure_aipe}

We start with defining the $\aipe$ problem as:

\begin{definition}[Adaptive Inner Product Estimation~($\aipe$)]
\label{def:AIPE}
Let $X=\{x_1,\ldots,x_n\}\in (\R^d)^n$ be a dataset of dimension $d$ on unit sphere and let $q\in \R^d$ be a query point on unit sphere. The \emph{Adaptive Inner Product Estimation} ($\aipe$) data structure, has the following guarantee: with probability at least $1-\delta$ we have 
\begin{align*}
    (1+\epsilon)\langle x_i,q\rangle-\epsilon \leq  w_i \leq (1-\epsilon)\langle x_i,q\rangle+\epsilon, ~~~ \forall i \in [n]
\end{align*}
where $w_i$ denotes the inner product estimation between $x_i$ and $q$.
\end{definition}

We provide a basic outline for this section: 
Section~\ref{sec:aipe_alg} introduces the details of our $\aipe$ algorithm.  Section~\ref{sec:aipe_result} presents our main results as a theoretical guarantees for our  $\aipe$ algorithm. Section~\ref{sec:aipe_ade} introduces our technique that reduces $\aipe$ to adaptive distance estimation problem.
Section~\ref{sec:aipe_ann} shows how to use our $\aipe$ algorithm for $\ann$. Section~\ref{sec:aipe_runtime} presents the running time analysis for our algorithm.

We define several notations to represent the time complexity for convenient,
\begin{definition}\label{def:time_function_T}
We fix parameter $\epsilon \in (0,1)$ and $\delta \in (0,1)$. 
We define ${\cal T}(x)$ as follows
\begin{align*}
    {\cal T}(x) := \wt{O}( \epsilon^{-2} x \log(1/\delta) ).
\end{align*}
\end{definition}

\subsection{Algorithm}\label{sec:aipe_alg}
This section proposes the $\aipe$ in Algorithm~\ref{alg:inner_product_estimation}. As shown in the algorithm, we have a data structure
that supports various adaptive inner product estimation operations. Firstly, we have an initialization function $\textsc{Init}$ that preprocesses the dataset into the data structure. Secondly, we could maintain our data structure with dynamic incremental insert and deletion operations. Thirdly, given a query $q$, we could estimate its distance with every element in the data structure. Finally, we also support an operation that outputs the item that has the maximum inner product with query approximately.

Our general idea is that the $\aipe$ problem can be reduced to adaptive distance estimation problem $\ade$ by transforms. Firstly, for vectors on the unit sphere, these two problems are equivalent to each other. Secondly, there exists efficient transforms shown in Section~\ref{sec:preli_transform} that map any vectors to unit vectors while preserving the order of pairwise inner product. As a result, we could perform $\aipe$ data structure shown in Algorithm~\ref{alg:inner_product_estimation} to accelerate the Frank-Wolfe algorithm.

We also would like to highlight that our data structure can support a sequence of adaptive queries. This capacity makes our idea possible to solve Frank-Wolfe efficiency problem as the weight queries are not independent to each other. We use a standard quantization technique that transforms the adaptive queries into independent vertices on the net. As a result, the data structure could handle adaptive queries sequence.

\begin{algorithm}[!ht]\caption{Adaptive Inner Product Estimation}\label{alg:inner_product_estimation}
\begin{algorithmic}[1]
\State {\bf data structure} \textsc{Adaptive Inner Product Estimation} \Comment{Theorem \ref{thm:aipe}}
\State {\bf members}
\State \hspace{4mm} \textsc{AdaptiveDistanceEstimation ADE}
\State {\bf end members}
\State 
\Procedure{Init}{$x_1,x_2,\cdots,x_n,\epsilon,\delta$} 
\Comment{Lemma~\ref{lem:aipe_init}}
    \State \textsc{ADE.Init}$(x_1,x_2,\cdots,x_n,\epsilon,\delta)$
\EndProcedure 
\State 
\Procedure{Insert}{$z \in \R^d$} 
\Comment{Lemma~\ref{lem:aipe_insert}}
    \State \textsc{ADE.Insert}$(z)$
\EndProcedure 
\State 
\Procedure{Delete}{$i\in [n]$}
\Comment{Lemma~\ref{lem:aipe_delete}}
    \State \textsc{ADE.Delete}$(i)$
\EndProcedure
\State
\Procedure{Query}{$q \in \R^d$} \Comment{Lemma~\ref{lem:query_AIPE} and Lemma~\ref{lem:aipe_query}}
    \State $d_1, d_2, \cdots, d_n \gets\textsc{ADE.Query}(q)$
    \For {$i=1,2,\cdots,n$}
        \State ${w}_i = 1-\frac{1}{2}{d}_i^2$
    \EndFor 
    \State \Return $\{{w}_i\}_{i=1}^n$
\EndProcedure 
\State 
\Procedure{QueryMax}{$q \in \R^d$} \Comment{Lemma~\ref{lem:query_max_AIPE} and Lemma~\ref{lem:aipe_query_max} }
    \State ${d}_1, {d}_2, \cdots, {d}_n \gets \textsc{ADE}.\textsc{Query}(q)$ 
    \State $i\gets \arg\min_{i\in [n]}~d_i$ 
    \State \Return $x_i$
\EndProcedure 
\State {\bf end data structure}
\end{algorithmic}
\end{algorithm}

\subsection{Main Result}\label{sec:aipe_result}

For simplicity, we only provide an informal statement in Theorem~\ref{thm:aipe_informal} which does not include the procedure \textsc{QueryMax}. This is because given the procedure \textsc{Query} that outputs a list of estimates, \textsc{QueryMax} is just taking the minimum over them. The smallest $\ell_2$ distance corresponds to the maximum (approximate) inner product, and hence facilitates enough functionality for our data structure to implement the Frank-Wolfe iteration. We give full details of the data structure in this and subsequent sections.

\begin{theorem}[Adaptive Inner Product Estimation, formal version of Theorem~\ref{thm:aipe_informal}]\label{thm:aipe}
Let $\T$ be defined as Definition~\ref{def:time_function_T}. 
There is a data structure requires $\T(nd)$ space for  Adaptive Inner Product Estimation Problem with the following procedures:
\begin{itemize}
    \item \textsc{Init}$(  \{x_1, x_2, \dots, x_n\}\subset \mathbb{S}^{d-1}, \epsilon \in (0,1), \delta \in (0,1))$: Given data points $\{x_1, x_2, \dots, x_n\}\subset \mathbb{S}^{d-1}$, an accuracy parameter $\epsilon$ and a failure probability $\delta$ as input, the data structure preprocesses in time $\T(nd)$.
    \item \textsc{Insert}$(z \in \R^d)$: Given a vector $z \in \R^d$, the data structure insert $z$ in time $\T(d)$.
    \item \textsc{Delete}$(i \in [n])$: Given an index $i \in [n]$, the data structure deletes $x_i$ in time $\T(d)$.
    \item \textsc{Query}$(q \in \mathbb{S}^{d-1})$: Given a query point $q \in \mathbb{S}^{d-1}$, the \textsc{Query} operation takes $q$ as input and approximately estimates the inner product of $q$ and all the data points $\{x_1, x_2, \dots, x_n\}\subset \mathbb{S}^{d-1}$ in time $\T(n+d)$ i.e. it provides a set of estimates $\{\tilde{w}_i\}_{i=1}^n$ such that:
    \begin{align*}
       \forall i \in[n], (1+\epsilon) \cdot \langle q,x_{i}\rangle - \epsilon \leq \tilde{w}_{i} \leq (1-\epsilon) \cdot \langle q,x_{i}\rangle + \epsilon
    \end{align*}
     with probability at least $1 -\delta$, even for a sequence of adaptively chosen queries. 
     \item \textsc{QueryMax}$(q\in\mathbb{S}^{d-1})$: Given a query point $q \in \mathbb{S}^{d-1}$, the \textsc{QueryMax} operation takes $q$ as input and solves the $(1+\epsilon, r)$-$\ann$ data structure problem, where $r\in (0,2)$ satisfies $\min_{x\in X}\|x-q\|_2\leq r$, in time $\T(n+d)$.  
\end{itemize}
\end{theorem}

\begin{remark}
The above theorem can be viewed as a variation of Theorem~\ref{thm:aipe} in \cite{sxz22}.
\end{remark}

\begin{proof}

For the proof of running time for operation\textsc{Init} see Lemma~\ref{lem:aipe_init}.

For the proof of running time for operation \textsc{Insert}, see Lemma~\ref{lem:aipe_insert}.

For the proof of running time for operation \textsc{Delete}, see Lemma~\ref{lem:aipe_delete}. 

For operation \textsc{Query}, the runtime follows from Lemma~\ref{lem:aipe_query}. Lemma~\ref{lem:query_AIPE} provides the proof of  correctness.

For operation \textsc{QueryMax}, the runtime is due to Lemma~\ref{lem:aipe_query_max}. We give the correctness proof in Lemma~\ref{lem:query_max_AIPE}.
\end{proof}

\subsection{From \texorpdfstring{$\ade$}{~} to \texorpdfstring{$\aipe$}{~}}\label{sec:aipe_ade}

We introduce the Adaptive Distance Estimation ($\ade$) data structure.

\begin{lemma}[Theorem 1.4 of~\cite{cn22}]\label{lem:cn22_ADE}
Let $\epsilon \in (0,0.1)$ be an accuracy parameter. Let $\delta\in (0,0.1)$ denote a failure probability. Let $\T$ be defined as Definition~\ref{def:time_function_T}. Then, there exists a data structure that uses $\T(nd)$ space and is initialized correctly with probability at least $1-\delta$ and supports the following operations:
\begin{itemize}
    \item \textsc{Init}$(x_1, \cdots, x_n)$. This operation takes $\T(nd)$ time.
    \item \textsc{Query}$(q \in \R^d)$ Output a vector $d \in \R^n$ such that 
    \begin{align*}
    (1-\epsilon)\|x_i-q\|_2 \leq d_i \leq (1+\epsilon)\|x_i-q\|_2,
    \end{align*}
    holds with probability at least $1-\delta$. This operation runs in $\T(n+d)$ time. The query can be chosen adaptively.
    \item \textsc{Update}$(i \in [n],z \in \R^d)$. Replace $x_i$ by $z$. This operation takes $\T(d)$ time.
\end{itemize}

\end{lemma}

Given an $\ade$ data structure, we would like to use it and solve the $\aipe$ problem. We use a reduction from previous work \cite{sxz22}.

\begin{lemma}[Special case of Lemma 6.5 in \cite{sxz22}]\label{lem:query_AIPE}
Let $X=\{x_1,\ldots,x_n\}\subset \R^{d-1}$ be the dataset on the unit sphere with $n$ points. Let $q\in \mathbb{S}^{d-1}$ be the unit query vector. The procedure \textsc{Query}$(q)$ in Algorithm~\ref{alg:inner_product_estimation} outputs a list of estimates $\{w_i\}_{i=1}^n$ such that
\begin{align*}
    (1+\epsilon)\langle x_i,q\rangle-\epsilon \leq w_i \leq (1-\epsilon)\langle x_i, q\rangle+\epsilon.
\end{align*}
\end{lemma}

\subsection{From \texorpdfstring{$\aipe$}{~} to \texorpdfstring{$\ann$}{~}}\label{sec:aipe_ann}

Before proceeding to the formal proof, we first show that suppose we can estimate all the inner product approximately, then we can solve the $\ann$ data structure problem.

\begin{lemma}
\label{lem:query_max_AIPE}
Let $X\subset \mathbb{S}^{d-1}$ be the dataset and let $q\in\mathbb{S}^{d-1}$ be a query point. Suppose for some $r\in (0,2)$ we have $\min_{x\in X} \|x-q\|_2\leq r$.
Then, \textsc{QueryMax} procedure of Algorithm~\ref{alg:inner_product_estimation} with precision parameter $\epsilon$ solves the $(1+\epsilon,r)$-$\ann$ problem.
\end{lemma}

\begin{proof}
Let $x\in X$ be the point that minimizes its distance with $q$, we have $\|x-q\|_2\leq r$ and let $d_x$, $d_y$ be the output given by the $\ade$ data structure on point $x$, $y$, respectively. Suppose for some $y\in X$ we have $d_y\leq d_x$ then
\begin{align*}
    d_y \leq & ~ d_x \\
    \leq & ~ (1+\epsilon)\|x-q\|_2 \\
    \leq & ~ (1+\epsilon) r.
\end{align*}

Further, we have that $\|y-q\|_2\leq \frac{d_y}{1-\epsilon}$, therefore, we conclude that
\begin{align*}
    \|y-q\|_2 \leq & ~ \frac{1+\epsilon}{1-\epsilon} r\\
    \leq & ~ (1+3\epsilon)r,
\end{align*}
where the second step is using the reason that $\epsilon \in (0,0.1)$.

Finally, if we re-scale $\epsilon$ by $\epsilon/3$, we get the desired result.
\end{proof}

\subsection{Running Time}\label{sec:aipe_runtime}
We prove the running time of various procedures of Algorithm~\ref{alg:inner_product_estimation}.

\begin{lemma}
\label{lem:aipe_init}
The procedure \textsc{Init} of Algorithm~\ref{alg:inner_product_estimation} takes time ${\cal T}(nd)$
\end{lemma}

\begin{proof}
Note that in \textsc{Init}, we simply initialize an $\ade$ data structure. By Lemma~\ref{lem:cn22_ADE}, it takes time ${\cal T}(md)$.
\end{proof}

\begin{lemma}
\label{lem:aipe_insert}
The procedure \textsc{Insert} of Algorithm~\ref{alg:inner_product_estimation} takes time $ {\cal T}(d) $.
\end{lemma}

\begin{proof}
Note that in \textsc{Insert}, we simply initialize an $\ade$ data structure. By Lemma~\ref{lem:cn22_ADE}, it takes time ${\cal T}(d)$.
\end{proof}

\begin{lemma}
\label{lem:aipe_delete}
The procedure \textsc{Delete} of Algorithm~\ref{alg:inner_product_estimation} takes time ${\cal T}(d)$.
\end{lemma}

\begin{proof}
Note that in \textsc{Delete}, we simply initialize an $\ade$ data structure. By Lemma~\ref{lem:cn22_ADE}, it takes time ${\cal T}(d)$.
\end{proof}

\begin{lemma}
\label{lem:aipe_query}
The procedure \textsc{Query} of Algorithm~\ref{alg:inner_product_estimation} takes time ${\cal T}(n+d)$.
\end{lemma}

\begin{proof}
Note that in \textsc{Query}, we first use \textsc{Query} of $\ade$ to output a list of estimates, then perform a simple transformation on all of them. By Lemma~\ref{lem:cn22_ADE}, it takes time ${\cal T}(n+d)$.
\end{proof}

\begin{lemma}
\label{lem:aipe_query_max}
The procedure \textsc{QueryMax} of Algorithm~\ref{alg:inner_product_estimation} takes time ${\cal T}(n+d)$.
\end{lemma}

\begin{proof}
Similar to \textsc{Query}, \textsc{QueryMax} first uses \textsc{Query} of $\ade$ to output a list of estimates, then finds the vector with the smallest estimate. By Lemma~\ref{lem:cn22_ADE}, it takes time ${\cal T}(n+d)$.
\end{proof}

\section{\texorpdfstring{{\lsh}-$\jl$}{~} Data Structure}\label{sec:lsh_jl_data_stucture}
In this section, we introduce the {\lsh}-JL data structure for efficient $\maxip$ in high dimension.

\subsection{Algorithm}
This section proposes {\lsh}-$\jl$ algorithm in Algorithm~\ref{alg:lsh_jl}. We start with setting our approximation multiplicative error to $\epsilon$ and failure probability to $\delta$. Next,
given a dataset with size $n$ and dimension $d$, our data structure contains several components: 
\begin{itemize} 
\item[1).] $k_{\mathrm{JL}}$ number of $\jl$
transform matrices, each is in $\R^{s\times d}$, where $s=O(\epsilon^{-2}\log(n/\delta))$. 
\item[2).] We use $k_{\mathrm{JL}}\cdot k_{\mathrm{LSH}}$ to represent the number of different {\lsh}
data structures.
\end{itemize}
Here we set $k_{\mathrm{JL}}=O((d+\log(1/\delta)) \cdot \log(nd))$.

We choose $k_{\mathrm{LSH}}=s\log(nd/\delta)$. 

In the $\textsc{Init}$ procedure, we first apply all the $\jl$ matrices on the dataset. As a result, we would get 
$k_{\mathrm{JL}}$ sketches for each dataset. Next, we use the sketches to build {\lsh} data structures. For each sketch, we build $k_{\mathrm{LSH}}$ {\lsh} data structures. In the $\textsc{QueryMax}$ procedure, we first randomly sample $l=O(\log(n/\delta))$ sketches with replacement from the $\jl$ matrices. Next, for each sampled matrix, we apply it on the query and use the query sketch to find maximum inner product items from the $k_{\mathrm{LSH}}$ {\lsh} data structures it belongs to.

Our algorithm contains two major techniques: 1).\ We combine the {\lsh} with $\jl$ transform to further reduce the time complexity on $d$. We remark that this combination is no-trivial as we need to provide clear guarantees for search quality, 2).\ we perform a random sub-sampling on the $\jl$ matrices. This random sampling reduce the query time complexity and further improves the overall running time efficiency.

\begin{algorithm}[!ht]\caption{{\lsh}-$\jl$ algorithm}\label{alg:lsh_jl}
\begin{algorithmic}[1]
\small
\State {\bf data structure} LSH-JL \Comment{Theorem~\ref{thm:robust_maxip}
}
\State {\bf members}
\State \hspace{4mm} $\epsilon\in (0,1)$
\State \hspace{4mm} $\delta\in (0,1)$
\State \hspace{4mm} $d\in \mathbb{N}_+$ \Comment{dimension}
\State \hspace{4mm} $n \in \mathbb{N}_+$ \Comment{size of dataset}
\State \hspace{4mm} $s\in \mathbb{N}_+$
\State \hspace{4mm} $k_{\mathrm{JL}} \in \mathbb{N}_+$ \Comment{number of $\jl$s to use}
\State \hspace{4mm} $k_{\mathrm{LSH}} \in \mathbb{N}_+$ \Comment{number of LSHs for each $\jl$}
\State \hspace{4mm}$S_1,\ldots,S_{k_{\mathrm{JL}}} \in \R^{s\times d}$
\State \hspace{4mm}$\mathrm{lsh}_1,\ldots,\mathrm{lsh}_{k_{\mathrm{JL}}\cdot k_{\mathrm{LSH}}}$
\State {\bf end members}
\State
\Procedure{Init}{$X=\{x_1,\ldots,x_n\},d, n, \epsilon, \delta, c$}
\State $\epsilon\gets \epsilon, \delta\gets \delta$
\State $d\gets d, n\gets n$
\State $k_{\mathrm{JL}}\gets O((d+\log(1/\delta))\log(nd))$
\State $s\gets O(\epsilon^{-2}\log(n/\delta))$
\State $k_{\mathrm{LSH}}\gets s\log(md/\delta)$
\State Initialize $S_1,\ldots,S_{k_{\mathrm{JL}}}\in \R^{s\times d}$ to be random Gaussian matrices
\State Initialize $\text{LSH}_1,\ldots,\text{LSH}_{k_{\mathrm{JL}}\cdot k_{\mathrm{LSH}}}$ to be locality-sensitive hashing data structures
\State ${\sf S}\gets \begin{bmatrix}
S_1^\top & S_2^\top & \ldots & S_{k_{\mathrm{JL}}}^\top
\end{bmatrix}^\top \in \R^{sk_{\mathrm{JL}}\times d}$
\State $Y\gets X{\sf S}^\top$
\For{$i=1\to k_{\mathrm{JL}}$}
\State $Y_i\gets Y_{*,\{(i-1)k_{\mathrm{JL}},\ldots,ik_{\mathrm{JL}} \}}$ \Comment{Sketched columns correspond to $i$-th JL}
\For{$j=1\to k_{\mathrm{LSH}}$}
\State $\text{LSH}_{(i-1)k_{\mathrm{LSH}}+j}\gets \text{LSH}.\textsc{Init}(Y_i, n, s, c)$
\EndFor
\EndFor
\EndProcedure
\State
\Procedure{QueryMax}{$q\in \mathbb{S}^{d-1}, \tau\in (0,1)$}
\State $l\gets O(\log(n/\delta))$
\State Sample $j_1,\ldots,j_l$ with replacement from $[k_{\mathrm{JL}}]$
\For{$i\in [l]$}
\State $\wt q\gets S_{j_i}q$
\For{$k=1\to k_{\mathrm{LSH}}$}
\State $y\gets \text{LSH}_{(j_i-1)k_{\mathrm{LSH}}+k}.\textsc{Query}(\wt q, \tau)$
\If{$\langle q, y\rangle \geq (1-\epsilon)c\tau$}
\State \Return $y$
\EndIf
\EndFor
\EndFor
\State \Return Fail
\EndProcedure
\end{algorithmic}
\end{algorithm}

\subsection{Adaptive Robust Johnson-Lindenstrauss Transform}
We first recall the Johnson-Lindenstrauss transform~\cite{jl84}:
\begin{definition}[Johnson-Lindenstrauss transform ($\mathsf{JLT}$)]
\label{def:JLT}
Let $\{x_1,\ldots,x_n\}\in (\R^d)^n$, we say a distribution $\Pi$ over $s\times d$ matrices is a $(n,\epsilon,\delta)$-${\sf JLT}$ if for any $S\sim \Pi$, we have
\begin{align*}
    \Pr[\|S(x_i-x_j)\|_2^2\in (1\pm\epsilon) \|x_i-x_j\|_2^2] \geq & ~ 1-\delta, ~~ \forall (i,j)\in [n]\times [n].
\end{align*}
\end{definition}

We remark that in order to obtain this property for all $n^2$ pairs of point, it suffices to obtain the following guarantee for any fixed point $x\in \R^d$:
\begin{align*}
    \Pr[\|Sx\|_2^2 \in (1\pm \epsilon) \|x\|_2^2] \geq & ~ 1-\delta,
\end{align*}

Next, we will apply union bound over $n^2$ events (where each event is a pair of two points), we are done.

Lemma~\ref{lem:robust_JL} shows that by using many independent Johnson-Lindenstrauss matrices.

\begin{lemma}[Lemma 7.12 in \cite{sxz22}]\label{lem:robust_JL}
We define $k_0:=(d+\log(1/\delta))\log(nd)$. 
Let $V:=\{v_1,\ldots,v_n\}\in (\R^d)^n$ denote a set of points. Let $\epsilon\in (0,1)$ denote an accuracy parameter. Let $\delta\in (0,1)$ denote a failure probability. Furthermore, let $\{S_i\}_{i=1}^k \subset \R^{b\times d}$ for $k\geq \Omega(k_0)$ such that each $S_i$ is an independent $(n+1,\epsilon,0.99)$-${\sf JLT}$ matrix~(Def. \ref{def:JLT}) with $\|S_i\|_F\leq d$. Then we have
\begin{align*}
    \forall q\in \mathbb{S}^{d-1}, \forall v\in V, \sum_{i=1}^k {\bf 1}[\|S_i(q-v)\|_2^2\in (1\pm O(\epsilon))\|q-v\|_2^2+\alpha]\geq 0.94k
\end{align*}
with probability at least $1-\delta$ and $\alpha\leq O(\frac{1}{(nd)^5})$.
\end{lemma}

\subsection{Solving \texorpdfstring{$\ann$}{~} using \texorpdfstring{{\lsh}}{~} }

In this section, we introduce the formal statement and the proofs that uses {\lsh} to solve  $(c,\tau)$-$\maxip$ problem for unit vectors.

\begin{corollary}[]\label{coro:maxip_lsh_formal}

Given a unit vectors set $Y$ with size $n$ and two parameters $c \in (0,1)$, $\tau \in(0,1)$, with success probability at least 0.9, we could solve $(c,\tau)$-$\maxip$ problem for any query unit vector $x$ with respect to $Y$ in $O(d\cdot n^{\rho})$ query time, 
$O(n^{1+o(1)} + d n)$ space and $O(dn^{1+o(1)})$ preprocessing time. Here $\rho:=  \frac{2(1-\tau)^2}{(1-c\tau)^2}-\frac{(1-\tau)^4}{(1-c\tau)^4}+o(1)$.

\end{corollary}
\begin{remark}
The succeed probability of current statement is 0.99. For any $\delta $, we can decrease the failure probability to $\delta$ by using $\log(1/\delta)$ copies of data structure.
\end{remark}

\begin{proof}
It is sufficient to show that $\|a-b\|_2^2= 2 - 2\langle a , b\rangle$ for any $a,b\in  \mathbb{S}^{d-1}$. In this way,  if solves $(\ov{c}, r)$-$\ann$ via a  {\lsh} data-structure. This data structure could also solve $(c, \tau)$-$\maxip$ with parameter $\tau = 1-0.5 r^2$ and $c = \frac{1-0.5 \ov{c}^2 r^2}{1 - 0.5 r^2}$. Next, we write $\ov{c}^2$ as
\begin{align*}
\ov{c}^2= \frac{ 1 - c(1-0.5 r^2) }{0.5r^2} = \frac{1 - c \tau }{1-\tau} .
\end{align*}

Next, it suffices to show that if we initialize {\lsh} with Theorem~\ref{thm:ar17:formal}, we can solve $(c,\tau)$-$\maxip$ by a $(\ov{c}, r)$-$\ann$ data structure. Moreover,
the query time will be $O(d \cdot n^{\rho})$. The space complexity will be $O(n^{1+o(1)} + d n)$. The preprocessing time will be $O(dn^{1+o(1)})$. Here the $\rho$ is

\begin{align*}
    \rho = & ~ \frac{2}{\ov{c}^2} -\frac{1}{\ov{c}^4} +o(1) \\
    = & ~ \frac{2(1-\tau)^2}{(1-c\tau)^2}-\frac{(1-\tau)^4}{(1-c\tau)^4}+o(1). 
\end{align*}

Therefore, it completes the proof.

\end{proof}

We could also improves this $\rho$ with more preprocessing time as shown in ~\cite{ar15}.

Next, we apply Corollary~\ref{coro:maxip_lsh_formal} to the projected version of $\maxip$.

\begin{corollary}[]\label{coro:proj_maxip_lsh}
Let $\phi, \psi: \R^d \rightarrow \R^k$ be a pair of transformations. Moreover, suppose it takes ${\cal T}_{\phi}$ and ${\cal T}_{\psi}$ to compute $\phi(x)$ and $\psi(y)$, respectively. Given a dataset $Y$ of size $n$ so that $\psi(Y) \subset \mathbb{S}^{k-1}$ and two parameters $c, \tau \in(0,1)$, with success probability at least $0.9$, we can solve $(c,\phi,\psi,\tau)$-$\maxip$ with respect to $(x,Y)$,  where $x$ is a query vector and $\phi(x)\in \mathbb{S}^{k-1}$ in $O(d\cdot n^{\rho}+{\cal T}_{\phi})$ query time, $O(n^{1+o(1)} + d n)$ space and $O(dn^{1+o(1)}+{\cal T}_{\psi}n)$ preprocessing time. Here $\rho:=  \frac{2(1-\tau)^2}{(1-c\tau)^2}-\frac{(1-\tau)^4}{(1-c\tau)^4}+o(1)$.
\end{corollary}

\begin{proof}

We preprocess the dataset with two steps
\begin{itemize}
    \item Compute $\psi(y)$ for every $y\in Y$ requires $O(n{\cal T}_{\psi})$ time.
    \item  Preprocess $\psi(Y)$  into {\lsh} following Corollary~\ref{coro:maxip_lsh_formal} requires
     \begin{itemize}
         \item $O(n^{1+o(1)}+dn)$ space,
         \item $O(dn^{1+o(1)})$ time.
     \end{itemize}
     
\end{itemize}

We also perform query in two steps
\begin{itemize}
    \item  Compute $\phi(x)$ takes $O({\cal T}_{\phi})$ time.
    \item 
    The query time complexity of {\lsh} following Corollary~\ref{coro:maxip_lsh_formal} is $O(d\cdot n^{\rho})$.
\end{itemize}
 Therefore, we finish our proof.
\end{proof}

We summarize the statement in a theorem. 
\begin{theorem}[]\label{thm:maxip_lsh}
Let $c, \tau \in(0,1)$ be two parameters. Given a set of $Y \subset \mathbb{S}^{d-1}$ with size $n$, we could have a data structure that takes preprocessing time ${\cal T}_{\mathsf{init}}$ and space ${\cal S}_{\mathsf{space}}$. Given an query point $x \in \mathbb{S}^{d-1}$, it takes  $O(d\cdot n^{\rho})$ time for query and:
\begin{itemize}
    \item if $ \maxip(x,Y)\geq\tau $, then we output a vector in $Y$ which is a $(c,\tau)$-$\maxip$ for $(x,Y)$ with success probability at least $0.9$\footnote{One can boost success probability from constant to $\delta$ by using $\log(1/\delta)$ independent data structures.}, where $\rho:= f(c,\tau)+o(1)$.
    \item otherwise, we output {\fail}.
\end{itemize}
Further, 
\begin{itemize}
   \item If ${\cal T}_{\mathsf{init}} = O( d n^{1+\rho})$ and ${\cal S}_{\mathsf{space}} = O(n^{1+\rho} + d n)$, then $f(c,\tau) = \frac{1-\tau}{1-2c\tau+\tau}$.
    \item If ${\cal T}_{\mathsf{init}} = O( d n^{1+o(1)})$ and ${\cal S}_{\mathsf{space}} = O( n^{1+o(1)} + d n)$, then $f(c,\tau) = \frac{2(1-\tau)}{(1-c\tau)}-\frac{(1-\tau)^2}{(1-c\tau)^2}$.
\end{itemize}

\end{theorem}

\subsection{Robust \texorpdfstring{{\lsh}}{~} Supports Adaptive Queries}

In this section, we focus on robust data structures for $\maxip$. We first define a noise version of $\maxip$.

\begin{definition}[Relaxed approximate $\maxip$]\label{def:quantized_approximate_maxip}
Given an $n$-vector set $Y \subset \mathbb{S}^{d-1}$ and three parameters $c\in (0,1)$, $\tau\in(0,1)$ and $\lambda\geq 0$, the $(c,\tau,\lambda)$-{$\maxip$} problem is defined as: given a query  $x$ which is a unit vector and $\max_{y\in Y}\langle x , y \rangle \geq \tau$, we would like to find a 
$z \in Y$ with condition that $\langle x , z \rangle \geq c \cdot \max_{y\in Y}\langle x , y \rangle - \lambda$. 
\end{definition}

In this paper, we use a net-argument to make a data structure for $(c,\tau,\lambda)$-{$\maxip$} that is robust to adaptive queries. We start with building a net where each lattice has the largest diameter $2\lambda$. Then, we locate each adaptive query to a lattice and use the lattice central point as a new query to {\lsh}. We note that an additive error $\lambda$ would be introduced here. However, we could use net-argument to union bound the failure probability since now the modified queries are independent. We present this idea in a statement as below:

\begin{corollary}[A robust version of Corollary~\ref{coro:maxip_lsh_formal}]\label{coro:maxip_lsh_adaptive} Given two parameters $c\in (0,1)$, $\tau\in(0,1)$, and a dataset $Y$ on unit sphere, for an adaptive query sequence $X\subset \mathbb{S}^{d-1}$ with maximum diameter $D_X$, with failure probability at most $\delta$, we could solve  $(c,\tau,\lambda)$-$\maxip$ for every $x\in X$  using $O(dn^\rho\cdot\kappa)$ query time,  $O((n^{1+o(1)}+dn)\cdot\kappa)$ space and $O(dn^{1+o(1)}\cdot\kappa)$  preprocessing time , where $\rho=   \frac{2(1-\tau)^2}{(1-c\tau)^2}-\frac{(1-\tau)^4}{(1-c\tau)^4}+o(1)$ and $\kappa:= d\log (nd D_X / (\lambda \delta ) ) $.
\end{corollary}

\begin{proof}
Using a net argument, we show that the prob. that $\geq$ one query 
in $X$ fails can be transformed to the prob. that $\geq$ one net vertex fail. We define our net as $\hat{N}$. As a result, it is sufficient to bound the overall failure probability as:

\begin{align*}
    \Pr[\exists q\in \hat{N}~~~\textrm{s.t all } ~ (c,\tau,\lambda)\textsc{-}{\maxip}(q,Y)~ \mathsf{fail} ]=n \cdot (dD_X/\lambda)^d \cdot (1/10)^{\kappa}\leq \delta
\end{align*}

Here the second step follows the definition $\kappa:= d\log (nd D_X / (\lambda \delta ) ) $.

Next, we use the same time and space complexity from Corollary~\ref{coro:maxip_lsh_formal}. As a result, we finish the proof.

\end{proof}

In this paper, we also need a projected version of $\maxip$ with additive errors.

\begin{definition}[]\label{def:quant_proj_approximate_maxip}
Let $\phi, \psi: \R^d \rightarrow \R^k$ be a pair of transformations. Given an $n$-points dataset $Y \subset \R^d $  where $\psi(Y)$ transforms the dataset onto unit sphere as well as three parameters $c,\tau\in(0,1)$ and  $\lambda\geq 0$, we define the problem $(c,\phi, \psi,\tau,\lambda)$-{$\maxip$} as follows:
given a query $x\in \R^d$ such that $\phi(x)$ is on the unit sphere, and $\max_{y\in Y}\langle \phi(x) , \psi(y) \rangle \geq \tau-\lambda$, the goal is to return vector $z \in Y$ with the guarantee $\langle \phi(x) , \psi(z) \rangle \geq c \cdot (\phi, \psi)\text{-}\maxip (x,Y)$.
\end{definition}

Next, it is sufficient to extend Corollary~\ref{coro:maxip_lsh_adaptive} for a sequence of adaptive queries.
\begin{corollary}[]\label{coro:proj_maxip_lsh_adaptive}
Let $\phi, \psi: \R^d \rightarrow \R^k$ be a pair of transformations. Moreover, suppose it takes ${\cal T}_{\phi}$ time to compute $\phi(x)$ and ${\cal T}_{\psi}$ time to compute $\psi(x)$. Given an $n$-point dataset $Y \subset \R^d $ such that $\psi(Y)$ is on unit sphere, and three parameters $c\in (0,1)$, $\tau\in(0,1)$ and  $\lambda\geq 0$, for any query $x\in\R^d$ from a sequence $X\subset \R^d$ with diameter $D_X$ such that $\phi(x)$ is on the unit sphere. There exists a Monte-Carlo data structrue that can solve $(c,\phi,\psi,\tau,\lambda)$-$\maxip(x,Y)$ using $O(dn^{\rho}\cdot\kappa+{\cal T}_{\phi})$ query time,  $O((dn^{1+o(1)} + d n)\cdot\kappa)$ space and $O(dn^{1+o(1)}\cdot\kappa+{\cal T}_{\psi}n)$ preprocessing time where $\rho:=  \frac{2(1-\tau)^2}{(1-c\tau)^2}-\frac{(1-\tau)^4}{(1-c\tau)^4}+o(1)$ and $\kappa:= d\log (nd D_X / (\lambda \delta ) ) $. The data structure succeeds with probability at least $1-1/\poly(n)$.
\end{corollary}

\subsection{Put Things Together}

The goal of this section is to prove Theorem~\ref{thm:robust_maxip}.

We first define some parameters.
\begin{definition}\label{def:robust_maxip}
Let $c \in (0,1)$. Let $\tau \in(0,1)$. Let $\lambda\in (0,1)$. Let $\epsilon\in (0,1)$ denote an accuracy parameter. Let $\delta\in (0,1)$ denote a failure probability. We define the following additional parameters: 
\begin{itemize}
    \item $\alpha=O(\frac{1}{(nd)^9})$, the additive error by quantizing with a net (Lemma~\ref{lem:robust_JL});
    \item $s=O(\epsilon^{-2}\log n)$, the dimension of $\mathsf{JLT}$;
    \item $k=O((d+\log(1/\delta))\log(nd))$, numober of copies of ${\sf JLT}$'s, for adaptivity;
    \item $\kappa=s\log(ns/(\lambda\delta))$, number of copies of \lsh's for each ${\sf JLT}$;
    \item $\wt \lambda=O(\sqrt{\frac{1-c\tau}{1-\tau}})\cdot (\lambda+\alpha)$, the additive error of {\lsh} by the second layer of net.
\end{itemize}
\end{definition}

Next, we state our data structure result,
\begin{theorem}[]\label{thm:robust_maxip}
Let ${\cal T}_S(x)$ denote the time of applying $S$ to a vector $x\in \R^d$. Given a set of $n$-points $Y \subset \mathbb{S}^{d-1}$ on the unit sphere, there exists a data structure with preprocessing time ${\cal T}_{\mathsf{init}}$ and space ${\cal S}_{\mathsf{space}}\cdot \kappa\cdot k$  so that for every query $x \in \mathbb{S}^{d-1}$ in an adaptive sequence $X=\{x_1,\ldots,x_T\}$, the query time is $\wt O(s n^{\rho}\cdot\kappa+{\cal T}_S(x))$:
\begin{itemize}
    \item if $ \maxip(x,Y)\geq\tau $, then we output a vector in $Y$ which is a $(c,\tau,\wt \lambda)$-$\maxip$ with respect to $(x,Y)$, where $\rho:= f(c,\tau)+o(1)$.
    \item otherwise, we output {\fail}.
\end{itemize}

Further, 
\begin{itemize}
    \item If ${\cal T}_{\mathsf{init}} = O( s n^{1+\rho}\cdot \kappa\cdot k)+{\cal T}_S(Y)\cdot k$ and ${\cal S}_{\mathsf{space}} = O(n^{1+\rho} + s n)$, then $f(c,\tau) = \frac{(1-\tau)(1+\epsilon)^2}{2-2c\tau-(1-\tau)(1+\epsilon)^2}$.
    \item If ${\cal T}_{\mathsf{init}} = O( s n^{1+o(1)}\cdot \kappa\cdot k)+{\cal T}_S(Y)\cdot k$ and ${\cal S}_{\mathsf{space}} = O( n^{1+o(1)} + s n)$, then $f(c,\tau) = \frac{2(1-\tau)(1+\epsilon)^2}{1-c\tau}-\frac{(1-\tau)^2(1+\epsilon)^4}{(1-c\tau)^2}$.
\end{itemize}
The probability that all queries succeed is at least $1-\delta$.
\end{theorem}

\begin{proof}
We first use Lemma~\ref{lem:robust_JL} to initiate $k\geq \Omega((d+\log(1/\delta))\log(nd))$ different ${\sf JLT}$ matrices with parameters $(m+1,\epsilon,0.99)$. Then, for each ${\sf JLT}$ matrix $S_i\in \R^{s\times d}$, we run the quantization process on it. Specifically, this requires us to use $\kappa=s\log (ns/(\lambda\delta))$ independent $\ann$ data structures due to Corollary~\ref{coro:maxip_lsh_adaptive}.

Throughout the proof, we will condition on the event that there exists some $i\in [k]$ such that $S_i$ preserves the pair-wise distances between any query point and points in $X$. To simplify the notation, we use $S$ to denote the corresponding ${\sf JLT}$ matrix $S_i$.

We consider the following: given a query point $Sx\in \R^s$, we quantize it into a point $\wh x\in \R^s$, then we use $\wh x$ as our query. Let $Sy$ be the nearest neighbor of $\wh x$, the $\ann$ data structure will output a point $Sy'$ with the guarantee that $\|Sy'-\wh x\|_2\leq \ov c\cdot \|Sy-\wh x\|_2$. Towards the end, we wish to have a bound on the term $\|x-y'\|_2$ in terms of $\|x-y\|_2$.
\begin{align*}
    \|Sy'-Sx\|_2 = & ~ \|Sy'-\wh x+\wh x-Sx\|_2 \\
    \leq & ~ \|Sy'-\wh x\|_2+\|\wh x-Sx\|_2 \\
    \leq & ~ \ov c\cdot \|Sy-\wh x\|_2+\lambda \\
    = & ~ \ov c\cdot \|Sy-Sx+Sx-\wh x\|_2+\lambda \\
    \leq & ~ \ov c\cdot (\|Sy-Sx\|_2+\lambda)+\lambda \\
    \leq & ~ \ov c\cdot ((1+\epsilon)\|y-x\|_2+\alpha+\lambda)+\lambda,
\end{align*}
the second step is by is by triangle inequality, the third step is by $\|Sy'-\wh x\|_2\leq \ov c\cdot \|Sy-\wh x\|_2$ and $\|\wh x-Sx\|_2\leq \lambda$. The fifth step is by again triangle inequality, and the last step is by $\|Sy-Sx\|_2\leq (1+\epsilon)\|y-x\|_2+\alpha$. 

On the other hand, we know that $\|x-y'\|_2 \leq \frac{\|Sy'-Sx\|_2+\alpha}{1-\epsilon}$, we hence conclude that
\begin{align*}
    \|x-y'\|_2 \leq & ~ \frac{\ov c\cdot (1+\epsilon)\|x-y\|_2+(1+\ov c)\lambda+(\ov c+1)\alpha}{1-\epsilon} \\
    = & ~ \underbrace{\ov c\cdot (1+O(\epsilon))}_{\wt c}\|x-y\|_2+\underbrace{(1+O(\epsilon))\cdot ((1+\ov c)\cdot \lambda+(\ov c+1)\cdot \alpha)}_{\wt \lambda}.
\end{align*}
By further setting $\wt r=\frac{r}{1+\epsilon}$, we conclude we get a $(\wt c,\wt r)$-$\ann$ data structure with additive error $\wt \lambda$. It is then instructive to compute the relationship between $\wt c,\wt r$ and $c,\tau$. Using the same proof strategy as in Theorem~\ref{thm:maxip_lsh}, we conclude a similar correctness result with the parameters $(c,\tau)$ set according to $\wt c$ and $\wt r$. 

Per Theorem~\ref{thm:maxip_lsh}, we have $\wt r^2=2-2\tau$, hence $\tau=1-0.5\wt r^2$ and $c=\frac{1-0.5\wt c\wt r^2}{1-0.5\wt r^2}$, which means $\wt c^2=\frac{1-c\tau}{1-\tau}$. Finally, use the relationship $\wt c=\ov c(1+\epsilon)$, we conclude that $\ov c^2=\frac{1-c\tau}{(1-\tau)(1+\epsilon)^2}$.

We can then consider using two different $\ann$ data structures:

{\bf Part 1.} If we are to use Theorem~\ref{thm:ar15:formal}, we have the following relationship between $\rho$ and $\ov c$: $\rho=\frac{1}{2\ov c^2-1}+o(1)$, this implies
\begin{align*}
    \frac{1}{2\ov c^2-1} = \frac{1}{2\frac{1-c\tau}{(1-\tau)(1+\epsilon)^2}-1}=\frac{(1-\tau)(1+\epsilon)^2}{2-2c\tau-(1-\tau)(1+\epsilon)^2}.
\end{align*}
Setting $f(c,\tau)=\frac{(1-\tau)(1+\epsilon)^2}{2-2c\tau-(1-\tau)(1+\epsilon)^2}$, we are done.

{\bf Part 2.} Using Theorem~\ref{thm:ar17:formal}, we know that $\rho=\frac{2}{\ov c^2}-\frac{1}{\ov c^4}+o(1)$, hence
\begin{align*}
    \frac{2}{\ov c^2}-\frac{1}{\ov c^4}=\frac{2(1-\tau)(1+\epsilon)^2}{1-c\tau}-\frac{(1-\tau)^2(1+\epsilon)^4}{(1-c\tau)^2}
\end{align*}
It suffices to set $f(c,\tau)$ to this value.

Use the relationship $\ov c^2=\frac{1-c\tau}{(1-\tau)(1+O(\epsilon))^2}$ we derived above, we can further simplify $\wt \lambda$:
\begin{align*}
    (1+O(\epsilon))\cdot ((1+\ov c)\cdot \lambda+\ov c\cdot \alpha) \leq & ~ O(1)
    \cdot \sqrt{\frac{1-c\tau}{1-\tau}}\cdot (\lambda+\alpha).
\end{align*}
Therefore, we simplify $\wt \lambda\leq O(\sqrt{\frac{1-c\tau}{1-\tau}}\cdot (\lambda+\alpha))$, we conclude that we get a $(c,\tau,\wt \lambda)$-$\maxip$.

We remark that $\wt\lambda$ merely affects the \emph{quality} of the $\maxip$ estimates, the success probability is still related to $\lambda$, since the quantization process generates a $\lambda$-net and it suffices to union bound over all vertices on this $\lambda$-net. Hence the number of independent data structures we need to use in order to boost success probability is $O(s\log (ns/(\lambda\delta)))$, as desired.

Finally, at query time, it suffices for us to sample $\Omega(\log(1/\delta))$ sketches uniformly at random to guarantee that with probability at least $1-2\delta$, there exists one sketch that preserves the distance of all query points. Hence, the query time is only blown up by a $\log(1/\delta)$ factor.

Therefore, we finish the proof.
\end{proof}

\section{Algorithms and Convergence Analysis}\label{sec:converge}

This section performs the algorithms and a convergence rate analysis for two algorithms that implement the $\FW$ process. We show that the two proposed algorithms significantly improves the running time efficiency of $\FW$ algorithm. We organize this section as below: We start with a summary of our results (Section~\ref{sec:converge_summary}). Next, we introduce the convergence rate of our accelerated $\FW$ algorithm in Section~\ref{sec:sublinear_fw}. Finally, we provide the discussion on the performance of our algorithm. 

\subsection{Algorithms and Summary}\label{sec:converge_summary}

In this section, we start with two algorithms implemented by data structures in Section~\ref{sec:data_structure_aipe} and~\ref{sec:lsh_jl_data_stucture}, 
then compare our methods with both original $\FW$ algorithm and the algorithm in \cite{xss21} in Table ~\ref{tab:slow_compare}. We include a reference to the statements and algorithms in detail.

\begin{algorithm}[!ht]\caption{Accelerated Frank-Wolfe algorithm, Formal version of Algorithm~\ref{alg:frank_wolfe_jl_informal} }\label{alg:frank_wolfe_jl_formal}
\begin{algorithmic}[1]
\Procedure{Accelerate-Frank-Wolfe}{$S \subset \R^d$,  $\tau\in(0,1)$, $c\in (0,1)$,  $k\in \mathbb{N} $, $m\in \mathbb{N}$, $d\in \mathbb{N}$, $n \in \mathbb{N}$} \Comment{Part I of Theorem~\ref{thm:frank_wolfe_aipe_formal}} 
\State Build $\phi, \psi : \R^d \rightarrow \R^{d+3}$ be two transforms as in Section~\ref{sec:preli_transform}
\State For $j\in[k]$, let $R_j: \R^{d+3}\rightarrow \R^m$ denote independent $\mathsf{JLT}$

\State {\bf static} \textsc{LSH} $\textsc{lsh}_1,\cdots,\textsc{lsh}_k$
\For{{ $j=1 \to k$}}
\State $\textsc{lsh}_j$.\textsc{Init}($R_j(\psi(S)),n,d+3,c$)
\EndFor

\State Init $w^0 \in {\cal B}$.

\State $T\leftarrow O(\frac{\beta D_{\max}^2}{c^2\epsilon})$
\For{ $t=1$~to~$T-1$}
    \State Sample $j_1,\cdots, j_r$ with replacement from $[k]$
    \For{$i\in[r]$}
    \State $s^t \leftarrow \textsc{lsh}_i.\textsc{Query}(\phi(w^t),\tau)$ 
    \If{$\langle s^t, \phi(w^t)\rangle \geq c\tau$}
    \State \textbf{break}
    \EndIf
    \EndFor
    \State $w^{t+1} \leftarrow w^{t}+\eta_t(s^t-w^t)$ \Comment{$\eta_t$ is defined as in Definition~\ref{def:eta_t}.}
\EndFor
\State \Return $w^{T}$
\EndProcedure
\end{algorithmic}
\end{algorithm}

\begin{algorithm}[!ht]\caption{Accelerated Frank-Wolfe, A formal version of Algorithm~\ref{alg:frank_wolfe_aipe_informal}}\label{alg:frank_wolfe_aipe_formal}
\begin{algorithmic}[1]
\Procedure{Accelerate-Frank-Wolfe}{$S \subset \R^d$, $\tau\in(0,1)$, $c\in (0,1)$, $d\in \mathbb{N}$,  $n \in \mathbb{N}$} \Comment{Part II of Theorem~\ref{thm:frank_wolfe_aipe_formal}} 
\State Build $\phi, \psi : \R^d \rightarrow \R^{d+3}$ be two transforms as in Section~\ref{sec:preli_transform}
\State {\bf static} \textsc{AIPE} \textsc{aipe}
\State \textsc{aipe}.\textsc{Init}($\psi(S),n,d+3,c$)
\State Init $w^0 \in {\cal B}$.
\State $T\leftarrow O(\frac{\beta D_{\max}^2}{c^2\epsilon})$
 
\For{ $t=1$ to $T-1$}
    \State $s^t \leftarrow \textsc{aipe}.\textsc{Query}(\phi(w^t),\tau)$ 
    \State $w^{t+1} \leftarrow w^{t} + \eta_t \cdot( s^t-w^t)$ \Comment{$\eta_t$ is defined as in Definition~\ref{def:eta_t}.}
\EndFor
\State \Return $w^{T}$
\EndProcedure
\end{algorithmic}
\end{algorithm}

\begin{table}[ht]
    \centering
    \begin{tabular}{|l|l|l|l|l|l|} \hline
        {\bf Algorithms} & {\bf References/Statements} & {\bf Preprocessing Time} & {\bf \#Iterations} & {\bf Cost per iter}  \\ \hline
        Algorithm~\ref{alg:frank_wolfe_formal} & \cite{j13}& 0 & $\beta D_{\max}^2/\epsilon$ & $dn+{\cal T}_{g}$  \\ \hline
        &\cite{xss21} & $d^2n^{1 + o(1)}$ & $c^{-2} \beta D_{\max}^2 / \eps$ & $d n^{\rho_1} + \kappa_g$ \\ \hline
        Algorithm~\ref{alg:frank_wolfe_jl_formal} &Theorem~\ref{thm:frank_wolfe_aipe_formal}(Part I) & $\Tmat(d,d,n) + dn^{1+o(1)}$ & $c^{-2} \beta D_{\max}^2 / \eps$ & $d + n^{\rho_1} + {\cal T}_g$ \\ \hline
        Algorithm~\ref{alg:frank_wolfe_aipe_formal} &Theorem~\ref{thm:frank_wolfe_aipe_formal}(Part II) & $\rho \cdot dn$ & $c^{-2} \beta D_{\max}^2 / \eps$ & $\rho_2 \cdot ( d + n ) + {\cal T}_g$ \\ \hline
    \end{tabular}
    \caption{Improvements of our algorithm over $\FW$ algorithm baselines. For simplicity of representation, we ignore the big-Oh notation in the table. Here we write ${\cal T}_{g}$ as the gradient computation time for $g$. We use $c\in (0,1)$ to denote the approximation parameter. Let $T$ denote the number of iterations of our algorithm. We set $\kappa :=  \Theta ( \log(T/\delta) )$. We define $\delta \in (0,1)$ as the failure probability of our algorithm. $\rho_1 =\frac{2(1-\tau)^2}{(1-c\tau)^2}-\frac{(1-\tau)^4}{(1-c\tau)^4}+o(1)$ is a value introduced when we use {\lsh}, and $\rho_2 = \frac{(1-\frac{1}{\tau})^2}{c^2}$ is a value introduced when we use $\aipe$.
    }
    \label{tab:slow_compare}
\end{table}

\subsection{Convergence Rate and Runtime of Accelerated Frank-Wolfe Algorithm}\label{sec:sublinear_fw}

This section presents the theoretical results to improve $\FW$ via our algorithm.

\begin{theorem}[Convergence rate and runtime of accelerated $\FW$, a formal version of Theorem~\ref{thm:frank_wolfe_aipe_informal}]\label{thm:frank_wolfe_aipe_formal}
Let $g : \R^d \rightarrow \R$ be a convex function that is $\beta$-smooth. Let ${\cal T}_{g}$ denote the time to compute $\nabla g(x)$.
Let $\psi_1, \psi_2: \R^d \rightarrow \R^k$ be a pair of transforms.  Let  $S \subset \R^d$ be a set with size $n$. Let ${\cal B} \subset \R^d$ be $S$'s convex hull as defined in Definition~\ref{def:cvx_hull}. Suppose $\lambda\leq c^{-2}\epsilon/4$, for any error $\epsilon$, there is an algorithm with that uses ${\cal S}_{\mathrm{space}}$ spaces, takes ${\cal T}_{\mathrm{prep}}$  preprocessing time , takes $T = O( c^{-2} \beta D_{\max}^2 / \epsilon )$ iterations and ${\cal T}_{ \mathrm{cost} }$ iteration cost complexity, begins from a random initialization $w^0\in{\cal B}$, and returns a $w^T \in {\cal B}$ that satisfies:
\begin{align*}
   g(w^T) - \min_{w\in \cal{B}} g(w) \leq \epsilon, 
\end{align*}
holds with probability at least $1-1/\poly(n)$.

For the running time, we have two choices. For the first choice, we use $\jl$ and \lsh (Algorithm~\ref{alg:frank_wolfe_jl_formal})
\begin{itemize}
    \item ${\cal S}_{\mathrm{space}} = O( d n^{1+o(1)} )$
    \item ${\cal T}_{\mathrm{prep}} = O( \Tmat(d,d,n) + dn^{1+o(1)} )$
    \item ${\cal T}_{\mathrm{cost}} = O(d + n^{\rho} + {\cal T}_g)$
    \item $T = O( c^{-2} \beta D_{\max}^2 / \epsilon )$ 
    \item Note that $c$, $\tau$ and $\rho$ has the following connection  $\rho:=  \frac{2(1-\tau)^2}{(1-c\tau)^2}-\frac{(1-\tau)^4}{(1-c\tau)^4}+o(1)$.
\end{itemize}
For the second choice, we use $\aipe$.(Algorithm~\ref{alg:frank_wolfe_aipe_formal})
\begin{itemize}
    \item ${\cal S}_{\mathrm{space}} = \wt{O} (\rho \cdot dn )$
    \item ${\cal T}_{ \mathrm{prep} } = \wt{O}( \rho \cdot dn )$
    \item ${\cal T}_{\mathrm{cost}} = O( \rho \cdot ( d + n ) + {\cal T}_g)$
    \item  $T = O(  c^{-2} \beta D_{\max}^2 / \epsilon  )$ 
    \item Note that $c$, $\tau$ and $\rho$ has the following connection $\rho:= \frac{(1-\frac{1}{\tau})^2}{c^2}$.
\end{itemize}
\end{theorem}

\begin{proof}

For the convergence argument, it is similar to that of Corollary~\ref{coro:proj_maxip_lsh}.

We give the proof of running time.

\paragraph{Choice 1 Result.} We provide a proof for Choice 1's space requirement and running time.

{\bf Space.} Recall that Choice 1 first uses $\wt O(d)$ independent Johnson-Lindenstrauss matrices, and each matrix is associated with $\wt O(1)$ {\lsh} data structures. Given $d$-dimensional input data of $n$ points, we first compress their dimension into $\wt O(1)$ using Johnson-Lindenstrauss, then store them in downstream {\lsh} data structures. For \emph{one} JL matrix and its corresponding $\wt O(1)$ {\lsh}'s, it takes $\wt O(n^{1+o(1)})$ space. Since there are $\wt O(d)$ JL matrices in total, the total space consumption is $\wt O(dn^{1+o(1)})$.

{\bf Preprocessing time.} Note that we can batch the $\wt O(d)$ JL matrices together. Since each of them has dimension $\wt O(1)$, this can be treated as a matrix-matrix multiplication between a matrix that has size $\wt O(d)$ by $d$ and another matrix which has size $d$ by $n$. Multiplying these two matrices together, it will use $\wt O(\Tmat(d,d,n))$ time. Then, we feed $\wt O(1)$-dimensional data into $\wt O(d)$ independent {\lsh} data structures, by Theorem~\ref{thm:robust_maxip}, this takes $\wt O(dn^{1+o(1)})$ time.
  
Hence, the preprocessing takes $\wt O(\Tmat(d,d,n)+dn^{1+o(1)})$ time.
 
{\bf Iteration Cost.} Computing $\nabla g(w^t)$ requires ${\cal T}_{g}$ time. Moreover, it needs $O(d)$ time to perform $\phi(w^t)$ based on Eq~\eqref{eq:asym_trans_direct}.

To realize the query operation, we first sample $\wt O(1)$ JL matrices, apply them to the input point takes $\wt O(d)$ time. We then feed points of dimension $\wt O(1)$ into the $\wt O(1)$ downstream {\lsh} data structures, this takes only $\wt O(n^\rho)$ time to retrieve $s^t$. After receiving $s^t$, we need $O(d)$ time to compute $w^{t+1}$. 
  
We write the complexity as $\wt O(d+n^\rho+{\cal T}_g)$ for $\rho:=  \frac{2(1-r)^2}{(1-cr)^2}-\frac{(1-r)^4}{(1-cr)^4}+o(1)$.
  
\paragraph{Choice 2 Result.}
 
We provide a proof for Choice 2's space requirement and running time.

{\bf Space.} Using Theorem~\ref{thm:aipe} (the space part), we get the space as ${\cal S}_{\mathrm{space}} = \wt{O}(\eps^{-2} nd \log (1/\delta))$, where $\eps = \frac{c-1}{3 (1-1/r)}$ denotes the accuracy guarantee used in Theorem~\ref{thm:aipe}. Plugging the value of $\eps$ gives us the space as ${\cal S}_{\mathrm{space}} = \wt{O}( \rho \cdot nd \log (1/\delta))$, for $\rho:= \frac{(1-\frac{1}{r})^2}{(c-1)^2}$.

{\bf Preprocessing Time.} Using Theorem~\ref{thm:aipe} (the preprocessing procedure part), and plugging the value of $\eps =\frac{c-1}{3 (1-1/r)}$ as above, we have the preprocessing time $\wt{O}(\rho \cdot n d \log (1/\delta))$.

{\bf Iteration Cost.} Using Theorem~\ref{thm:aipe} (the query procedure part), and plugging the value of $\eps =\frac{c-1}{3 (1-1/r)} $ as above, we have the cost per iteration as $\wt{O}(\rho \cdot ( n + d ) \log (1/\delta))$. 
 
As a result, we finish the proof.
 \end{proof}

 \subsection{Discussion}
 Our statement in Theorem~\ref{thm:frank_wolfe_aipe_formal} introduces two significant improvements over both original Frank-Wolfe algorithm and the algorithm in~\cite{xss21}. We summarize the advantages of two choices as below: 1).\ Using $\jl$ and {\lsh}, the cost per iteration can be reduced to $\wt O(d+n^\rho)$, which reduces a $d$ multiplicative factor over \cite{xss21} and  benefits large scale optimization. 2).\ Using $\aipe$, we are able to achieve near linear time preprocessing in the data structure, while having a $\wt{O}(d+n)$ iteration cost. This saves a lot of preprocessing for large scale dataset and benefits online algorithms.
\section{Herding Algorithm}\label{sec:herding}
This section explains how to use our data structure to tackle the Herding. We start with defining the Herding problem and the accelerated algorithm
in Section~\ref{sec:herding_problem}. Next, Section~\ref{sec:herding_converge} presents the convergence rate of our accelerated Herding algorithm. Finally, we introduce a discussion in Section~\ref{sec:herding_discuss}.
 
\subsection{Herding Problem}\label{sec:herding_problem}
We would like to use our algorithm to solve the Herding problem: given a kernel function $\Phi: \R^d \rightarrow \R^k$ and a set ${\cal X}\subset \R^d$, we define an estimation $\mu$ as 
\begin{align}\label{eq:herding_target}
    \mu=\E_{x\sim p(x)}[\Phi(x)] 
\end{align}
where $p(x)$ is a distribution defined on 
${\cal X}\subset \R^d$.

The Herding problem can be formulated as below~\cite{cws12}: we would like to find $\{x_1, x_2, \cdots, x_T \}$ from ${\cal X}$ and minimize $\|\mu-\sum_{t=1}^{T} v_t \Phi(x_t)\|_2$. Here we set $v_t$ to be positive or zero. The Herding algorithm solves this problem via the following rule.

\begin{align}\label{eq:herding_orginal}
    x_{t+1} =\arg\min_{x\in{\cal X}} \langle -w_t , \Phi(x) \rangle\notag\\
    w_{t+1}=w_t+\mu-\Phi(x_{t+1}) 
\end{align}

It is sufficient to show that the update rule above is equivalent to a problem that can be solved by Frank-Wolfe algorithm~\cite{blo12}. We define the problem as:

\begin{problem}[Herding]\label{prob:herding_fw}
We would like to optimize:
\begin{align*}
     \min_{w\in {\cal B}(S)} \frac{1}{2} \|w-\mu\|_2^2
\end{align*}

Here we assume $S=\Phi({\cal X})\subset \R^d$ is a set with size $n$, we would like to find a $w$ in the convex hull of $S$, such that its $\ell_2$ distance with $\mu$ is minimized. The convex hull has a diameter $D_{\max}$ (see Definition~\ref{def:cvx_hull}).
\end{problem}

\begin{algorithm}[!ht]\caption{Accelerated Herding, Part I}\label{alg:herding_lsh_jl_formal}
\begin{algorithmic}[1]
\Procedure{Accelerated-Herding}{$S \subset \R^d$,  $\tau\in(0,1)$, $c\in (0,1)$,  $k\in \mathbb{N} $, $m\in \mathbb{N}$, $d\in \mathbb{N}$, $n \in \mathbb{N}$} \Comment{Part I of Theorem~\ref{thm:frank_wolfe_aipe_formal}} 
\State Build $\phi, \psi : \R^d \rightarrow \R^{d+3}$ as in Section~\ref{sec:preli_transform}
\State For $j\in[k]$, let $R_j: \R^{d+3}\rightarrow \R^m$ denote independent JL transforms 

\State {\bf static} \textsc{LSH} $\textsc{lsh}_1,\cdots,\textsc{lsh}_k$
\For{{ $j=1 \to k$}}
\State $\textsc{lsh}_j$.\textsc{Init}($R_j(\psi(S)),n,d+3,c$)
\EndFor

\State Init a $w^0 \in {\cal B}$. 

\State $T\leftarrow O(\frac{\beta D_{\max}^2}{c^2\epsilon})$

\For{ $t=1$ to $T-1$}

    \State Sample $j_1,\cdots, j_r$ with replacement from $[k]$
    \For{$i\in[r]$}
    \State $s^t \leftarrow \textsc{lsh}_i.\textsc{Query}(\phi(w^t),\tau)$ 
    \If{$\langle s^t, \phi(w^t)\rangle \geq c\tau$} 
    \State \textbf{break}
    \EndIf
    \EndFor
       \State $w^{t+1}  \leftarrow w^t+\eta_t(s^t-w^t)$ \Comment{$\eta_t$ is defined as in Definition~\ref{def:eta_t}.}
\EndFor
\State \Return $w^{T}$
\EndProcedure
\end{algorithmic}
\end{algorithm}

\begin{algorithm}[!ht]\caption{Accelerated Herding}\label{alg:herding_aipe_formal}
\begin{algorithmic}[1]
\Procedure{Accelerated-Herding}{$S \subset \R^d$,  $\tau\in(0,1)$, $c\in (0,1)$, $d\in \mathbb{N}$,  $n \in \mathbb{N}$} \Comment{Part II of Theorem~\ref{thm:frank_wolfe_aipe_formal}} 
\State Build $\phi, \psi : \R^d \rightarrow \R^{d+1}$ as in Section~\ref{sec:preli_transform}
\State {\bf static} \textsc{AIPE} \textsc{aipe}
\State \textsc{aipe}.\textsc{Init}($\psi(S),n,d+3,c$)
\State Init a $w^0 \in {\cal B}$. 

\State $T\leftarrow O(\frac{\beta D_{\max}^2}{c^2\epsilon})$

\For{ $t=1$ to $T-1$}
    \State $s^t \leftarrow \textsc{aipe}.\textsc{Query}(\phi(w^t),\tau)$ 
     \State $w^{t+1}  \leftarrow w^t+\eta_t(s^t-w^t)$ \Comment{$\eta_t$ is defined as in Definition~\ref{def:eta_t}.}
\EndFor
\State \Return $w^{T}$
\EndProcedure
\end{algorithmic}
\end{algorithm}

\subsection{Convergence rate and runtime of accelerated Herding algorithm}\label{sec:herding_converge}
We compare our algorithm with original $\FW$ and \cite{xss21} in Table~\ref{tab:herd_compare}. We include the reference to the statements and algorithms in the table. It is shown that we could further improves the running time of $\FW$ algorithm for Herding.

\begin{table}[ht]
    \centering
    \begin{tabular}{|l|l|l|l|l|l|} \hline
        {\bf Algorithms}&{\bf References/Statements} & {\bf Preprocessing Time} & {\bf \#Iterations} & {\bf Cost per iter}  \\ \hline
        & \cite{blo12}  & 0  & $D_{\max}^2/\epsilon$  & $dn$  \\ \hline
        &\cite{xss21}  & $dn^{1+o(1)} + d^2 n$  &  $c^2 D^2_{\max}/\epsilon$  & $  dn^{\rho} $ \\ \hline Algorithm~\ref{alg:herding_lsh_jl_formal}& Choice 1, Theorem~\ref{thm:herding_jl_formal}  &  $\Tmat(d, d, n)+d n^{1+o(1)}$ & $c^2 D^2_{\max}/\epsilon$ & $  d+n^{\rho} $ \\ \hline 
        Algorithm~\ref{alg:herding_aipe_formal} &Choice 2, Theorem~\ref{thm:herding_jl_formal} & $dn$ & $c^2 D^2_{\max}/\epsilon$  & $  d+n $ \\
        \hline
    \end{tabular}
    \caption{Comparison between our algorithms and prior algorithms.}
    \label{tab:herd_compare}
\end{table}

Next, we show that our objective function is $1$-smooth.

\begin{lemma}[Lemma E.2 in \cite{xss21}]\label{lemma:1smooth}
$g(w)=\frac{1}{2}\|w-\mu\|_2^2$ a smooth function with smoothness $1$. Morover $g(w)$ is a convex function.
\end{lemma}

\begin{theorem}[Convergence result and runtime of accelerated Herding]\label{thm:herding_jl_formal}
Let ${\cal X}\subset \R^d$ denote a dataset and $\Phi: \R^d \rightarrow \R^k$ denote a linearized kernel transform. Let ${\cal B}(\Phi({\cal X}))$'s maximum diameter be $D_{\max}$. For a distribution $P$ defined on ${\cal X}$, we write $\mu= \E_{x\sim P}[\Phi(x)]$. Let $\rho\in(0,1)$, for any error parameter $\epsilon$, we have an iterative algorithm that uses ${\cal S}_{\mathrm{space}}$ space and ${\cal T}_{\mathrm{prep}}$ time in pre-processing, and ${\cal T}_{\mathrm{cost}}$ computation in each iteration, begins with a random initialization weight $w_0\in {\cal B}(\Phi({\cal X}))$,
and generates $w_T \in \R^k$ from ${\cal B}(\Phi({\cal X}))$ after $T$ iterations such that
\begin{align*}
   \frac{1}{2} \|w_T-\mu\|_2^2  \leq \min_{w\in \cal{B}} \frac{1}{2} \|w-\mu\|_2^2+\epsilon, 
\end{align*}
holds with probability at least $1-1/\poly(n)$.

Moreover, we could solve the problem with:
\begin{itemize}
    \item ${\cal S}_{\mathrm{space}} = O( dn^{1+o(1)})$, 
    ${\cal T}_{\mathrm{prep}} = O({\cal T}_{\mathrm{mat}} (d,d,n) + d n ^{1 + o(1)})$, 
     $T = O(D_{\max}^2 / \epsilon)$, \\ 
     and ${\cal T}_{\mathrm{cost}} = O( dn^{\rho} )$ using Algorithm~\ref{alg:herding_lsh_jl_formal}.
    \item${\cal S}_{\mathrm{space}} = O( dn)$ 
    , ${\cal T}_{\mathrm{prep}} = O(dn)$
    , $T = O({D_{\max}^2}/{\epsilon})$,
    \\ 
    and ${\cal T}_{\mathrm{cost}} = O( d+n )$, using Algorithm~\ref{alg:herding_aipe_formal}.
\end{itemize}
\end{theorem}

\begin{proof}
The key is to observe the function $g(\omega)=\frac{1}{2}\|w^T-\mu\|_2^2$ is a smooth function with smoothness parameter being 1 due to Lemma~\ref{lemma:1smooth}, and the gradient can be easily computed in time $O(d)$.

Next, by calling Theorem~\ref{thm:frank_wolfe_aipe_formal} with $\beta=1$ and ${\cal T}_g=d$, we could prove the our theorem.
\end{proof}

\subsection{Discussion}\label{sec:herding_discuss}
With our analysis, we show that our accelerated $\FW$ algorithm in  Algorithm~\ref{alg:herding_lsh_jl_formal} and Algorithm~\ref{alg:herding_aipe_formal} improves the Herding algorithm of~\cite{xss21}. These results would benefit the potential application of Herding algorithm in recommendation datasets that features are multiplied as an data augmentation.

\bibliographystyle{alpha}
\bibliography{ref}
\end{document}